\documentclass[acmsmall]{acmart}

\usepackage{pifont}
\usepackage{tikz}
\usepackage{algorithm}
\usepackage{multirow}
\usepackage[noend]{algpseudocode}
\usepackage{xspace}
\usepackage{amsmath}
\DeclareMathOperator*{\argmax}{arg\,max}
\usepackage{array}
\usepackage{graphicx}
\usepackage{subcaption}
\usepackage{listings}
\usepackage{adjustbox}
\usepackage{longfbox}
\usepackage{float}
\usepackage[dvipsnames]{xcolor}
\usepackage{booktabs}
\usepackage{placeins}
\renewcommand{\sectionautorefname}{\S\kern-2pt}

\algrenewcommand\algorithmicrequire{\textbf{Input:}}
\algrenewcommand\algorithmicensure{\textbf{Output:}}

\setcopyright{cc}
\setcctype{by-nc-nd}
\acmDOI{10.1145/3798257}
\acmYear{2026}
\acmJournal{PACMPL}
\acmVolume{10}
\acmNumber{OOPSLA1}
\acmArticle{149}
\acmMonth{4}
\received{2025-10-10}
\received[accepted]{2026-02-17}

\begin{document}
%
\title{\sys: Randomized Corpus Reduction for Fuzzing Seed Scheduling}

\author{Yuchong Xie}
\authornote{Equal Contribution.}
\orcid{0009-0008-0436-8183}
\affiliation{%
  \institution{Hong Kong University of Science and Technology}
  \city{Hong Kong}
  \country{China}
}
\affiliation{%
  \institution{Fudan University}
  \city{Shanghai}
  \country{China}
}
\email{yxiece@cse.ust.hk}

\author{Kaikai Zhang}
\authornotemark[1]
\orcid{0009-0002-9784-6477}
\affiliation{%
  \institution{Hong Kong University of Science and Technology}
  \city{Hong Kong}
  \country{China}
}
\email{kzhangcq@cse.ust.hk}

\author{Yu Liu}
\orcid{0000-0002-2207-1569}
\affiliation{%
  \institution{Fudan University}
  \city{Shanghai}
  \country{China}
}
\email{yuliu21@m.fudan.edu.cn}

\author{Rundong Yang}
\orcid{0009-0003-6567-7741}
\affiliation{%
  \institution{Fudan University}
  \city{Shanghai}
  \country{China}
}
\email{rundongyang22@m.fudan.edu.cn}

\author{Ping Chen}
\orcid{0000-0002-8517-0580}
\affiliation{%
  \institution{Fudan University}
  \city{Shanghai}
  \country{China}
}
\email{pchen@fudan.edu.cn}

\author{Shuai Wang}
\orcid{0000-0002-0866-0308}
\affiliation{%
  \institution{Hong Kong University of Science and Technology}
  \city{Hong Kong}
  \country{China}
}
\email{shuaiw@cse.ust.hk}

\author{Dongdong She}
\authornote{Corresponding author.}
\orcid{0000-0001-6655-0468}
\affiliation{%
  \institution{Hong Kong University of Science and Technology}
  \city{Hong Kong}
  \country{China}
}
\email{dongdong@cse.ust.hk}

\newcommand{\todo}[1]{\textbf{\textit{\textcolor{red}{#1}}}}
\newcommand{\fixme}[1]{\textcolor{blue}{#1}}
\newcommand{\sys}{RandSet\xspace}
\newcommand{\sysaflpp}{MinSet-A\xspace}
\newcommand{\syslibafl}{MinSet-L\xspace}
\newcommand{\syscentipede}{MinSet-C\xspace}
\newcommand{\aflpp}{AFL++\xspace}
\newcommand{\libafl}{LibAFL\xspace}
\newcommand{\centipede}{Centipede\xspace}

\newcommand{\sysgreedy}{RandSet-Greedy\xspace}
\newcommand{\sysgreedyshort}{R-G\xspace}
\newcommand{\sysnofma}{RandSet-Prioritize\xspace}
\newcommand{\sysnofmashort}{R-P\xspace}
\newcommand{\sysedgefeature}{RandSet-Edge\xspace}
\newcommand{\sysedgefeatureshort}{R-E\xspace}
\newcommand{\aflppcmin}{AFLPP-CM\xspace}
\newcommand{\aflppcq}{AFLPP-CQ\xspace}
\newcommand{\aflppminset}{AFLPP-MS\xspace}
\newcommand{\aflpprr}{AFLPP-RR\xspace}
\newcommand{\aflppsr}{AFLPP-SR\xspace}

\newcommand*\emptycirc[1][1ex]{\tikz[baseline=-0.5ex]\draw (0,0) circle (#1);} 

\newcommand*\halfcirc[1][1ex]{%
    \begin{tikzpicture}[baseline=-0.5ex]
    \draw[fill] (0,0)-- (90:#1) arc (90:270:#1) -- cycle ;
    \draw (0,0) circle (#1);
    \end{tikzpicture}}
    
\newcommand*\fullcirc[1][1ex]{\tikz[baseline=-0.5ex]\fill (0,0) circle (#1);}

\newcommand{\maxmagma}{7\xspace}
\newcommand{\uniquebug}{5\xspace}
\newcommand{\standsubsetratio}{4.03\%\xspace}
\newcommand{\fuzzbenchsubsetratio}{5.99\%\xspace}
\newcommand{\aflppstandsubsetratio}{13.40\%\xspace}
\newcommand{\aflppfuzzbenchsubsetratio}{15.72\%\xspace}
\newcommand{\maxstandalonecov}{16.58\%\xspace}
\newcommand{\maxaflppfuzzbenchcov}{3.57\%\xspace}
\newcommand{\maxlibaflfuzzbenchcov}{3.64\%\xspace}
\newcommand{\maxcentipedefuzzbenchcov}{7.72\%\xspace}
\newcommand{\standaloneover}{0.04\%\xspace}
\newcommand{\aflppfuzzbenchover}{1.17\%\xspace}
\newcommand{\libaflfuzzbenchover}{9.20\xspace}
\newcommand{\centipedefuzzbenchover}{0.35\xspace}
\newcommand{\standalonedistill}{4.00\%\xspace}
\newcommand{\aflppfuzzbenchdistill}{5.99\%\xspace}
\newcommand{\libaflfuzzbenchdistill}{17.27\xspace}
\newcommand{\centipedefuzzbenchdistill}{2.30\xspace}
\newcommand{\aflppfuzzbenchgreedy}{3.21\%\xspace}
\newcommand{\libaflfuzzbenchgreedy}{3.98\%\xspace}
\newcommand{\centipedefuzzbenchgreedy}{5.20\%\xspace}
\newcommand{\aflppfuzzbenchrandom}{1.10\%\xspace}
\newcommand{\libaflfuzzbenchrandom}{3.05\%\xspace}
\newcommand{\centipedefuzzbenchrandom}{5.61\%\xspace}
\newcommand{\numberwinstandalone}{14\xspace}
\newcommand{\numprograms}{35\xspace}
\newcommand{\numstandalone}{15\xspace} 
\newcommand{\numfuzzbench}{19\xspace} 
\newcommand{\numfuzzbenchaflpp}{19\xspace} 
\newcommand{\numfuzzbenchlibafl}{16\xspace} 
\newcommand{\numfuzzbenchcentipede}{18\xspace} 

\newcommand{\added}[1]{{#1}}
\newcommand{\replaced}[2]{{#2}}
\begin{abstract}
Seed explosion is a fundamental problem in fuzzing seed scheduling. It occurs when a fuzzer maintains a corpus with a huge number of seeds and fails to choose a promising one. Existing seed scheduling works focus on seed prioritization but suffer from the seed explosion since the seed corpus size is still huge. We tackle seed explosion from a new perspective, corpus reduction, i.e., compute a seed corpus subset.  
Corpus reduction can eliminate redundant seeds in the corpus and significantly reduce corpus size. However, this could lead to poor diversity in seed selection and severely impact the fuzzing performance. Meanwhile, effective corpus reduction incurs large runtime overhead.
In practice, it's challenging to adopt corpus reduction in fuzzing seed scheduling. Prior techniques like \texttt{cull\_queue}, AFL-Cmin and MinSet all suffer from poor seed diversity. AFL-Cmin and MinSet incur prohibitive runtime overhead and are hence only applicable to one-time task initial seed selection rather than high-frequency seed scheduling. 

%
We propose a novel randomized corpus reduction technique, RandSet, that can reduce the corpus size and yield diverse seed selection simultaneously. Meanwhile, the runtime overhead of RandSet is minimal, suiting a high-frequency seed scheduling task. Our key insight is to introduce randomness into corpus reduction so as to enjoy the two benefits of a randomized
algorithm: randomized output (i.e., diverse seed selection) and low runtime overhead.
Specifically, we formulate the corpus reduction in seed scheduling as a classic set cover problem and compute a randomized subset of seed corpus as a set cover to cover all features of the entire corpus. We then develop a novel seed scheduling approach using the randomized corpus subset. 
Our technique can effectively mitigate seed explosion by scheduling a small and randomized subset of the corpus rather than the entire corpus. 

We implement \sys on three popular fuzzers: AFL++, LibAFL and Centipede to showcase its general algorithmic design. We perform a comprehensive evaluation of \sys on three benchmarks: standalone programs, FuzzBench and Magma. Our evaluation results show that \sys can achieve significantly more diverse seed selection compared with other corpus reduction techniques.  \sys also yields high reduction ratio, achieving an average subset ratio of \standsubsetratio and \fuzzbenchsubsetratio after corpus reduction in terms of standalone programs and FuzzBench programs.
In terms of fuzzing performance gain from our randomized corpus reduction, \sys achieves a \maxstandalonecov gain on standalone programs and up to \maxaflppfuzzbenchcov gain on FuzzBench programs in AFL++. \sys triggers up to \maxmagma more ground-truth bugs than the state-of-the-art fuzzer on Magma, while introducing only 3.93\% overhead on standalone programs and as low as 1.17\% overhead on FuzzBench.

\end{abstract}




\maketitle

%

\section{Introduction}
Software fuzzing has emerged as one of the most effective and widely adopted techniques \cite{schumilo2017kafl, redqueen, she2019neuzz, liu2023vd,xie2025hfuzz,li2025enhancing, chen2022sfuzz, bao2025alarms, jiang2024binpre, sun2023property, zhang2025your, wong2025extraction,xiao2023phyfu} for discovering vulnerabilities and bugs in real-world programs. 
Seed scheduling is a crucial component in fuzzing that influences its performance~\cite{seed_select}. Seed scheduling periodically selects a seed from seed corpus at each scheduling round. A fundamental problem in seed scheduling is \emph{seed explosion}. It represents the \emph{monotonic} increasing of the seed corpus during the \emph{long-running} fuzzing campaigns. As fuzzing campaigns typically run for long periods (e.g., 24 hours in academic experiments \cite{klees2018evaluating} or continuous settings in OSS-Fuzz \cite{ossfuzz}), the corpus grows drastically. The continuous expansion of the seed corpus makes it increasingly difficult for a fuzzer to select a promising seed, as it must maintain and evaluate the huge seed corpus.
Previous works have proposed various seed scheduling algorithms that assign weights to each seed using different metrics to guide the seed selection~\cite{ecofuzz,lemieux2018fairfuzz, aflfast, slowfuzz, wang2021reinforcement, She2022EffectiveSS, Wang2020NotAC, zhang2022path}. 
However, we believe that these prioritization strategies still suffer from the seed explosion issue, primarily due to the fact that the seed corpus size does not change, and the fuzzer faces a massive number of choices. 

\replaced{A better solution}{A complementary solution} is corpus reduction~\cite{aflcullqueue, aflcmin, rebert2014optimizing, peachminset} i.e., computing a small subset of the huge seed corpus. However, it is quite challenging to adopt this technique in a real-world fuzzing campaign. First, corpus reduction has an inherent limitation --- poor diversity in seed selection, which poses a significant threat to the fuzzing performance. Fuzzing essentially is a random search task whose performance is highly sensitive to the diversity of seed selection~\cite{seed_select}. Hence, the limited seed choices caused by corpus reduction could severely affect fuzzing performance. Second, effective corpus reduction is a hard problem, given that the corpus subset should ideally preserve all the unique program features of the entire seed corpus. Prior work~\cite{rebert2014optimizing} shows that computing a minimal corpus subset that satisfies this requirement is an NP-hard problem. Fuzzing practitioners and researchers usually use a polynomial-time approximation algorithm to obtain a subset of the seed corpus. Third, the runtime cost of the corpus reduction is high. Seed scheduling is a high-frequency task that is invoked a substantial number of times during fuzzing. Meanwhile, each seed scheduling round requires one-time corpus reduction to mitigate the potential seed explosion issue, incurring prohibitive runtime overhead.  

We investigated existing corpus reduction techniques \cite{rebert2014optimizing, aflcmin, aflcullqueue, peachminset} and found that, unfortunately, there is no practical solution to applying corpus reduction for seed scheduling task. A widely-used corpus reduction technique is the \texttt{cull\_queue} function implemented in AFL~\cite{afl}. It proposes an efficient approximation technique for computing a subset of the seed corpus. Nevertheless, we observe that the seed corpus subset generated by \texttt{cull\_queue} suffers from poor diversity in seed selection. Even more surprisingly, our empirical study shows that AFL++ skips the \texttt{cull\_queue} strategy and abandons the corpus subset in over 98.39\% of scheduling rounds on FuzzBench programs and over 89.84\% on standalone programs. Instead, it simply performs weighted random sampling from the entire seed corpus. Consequently, this means that AFL++ suffers from seed explosion in the vast majority of fuzzing processes. Another line of corpus reduction techniques, such as MinSet and AFL-Cmin~\cite{rebert2014optimizing, aflcmin, peachminset}, are typically used in initial seed selection (\emph{one-time} task at the beginning of the fuzzing campaign) rather than in seed scheduling (\emph{periodic} task during the entire fuzzing campaign). They can reduce the size of seed corpus but at the cost of high runtime overhead, making them suitable only for initial corpus preparation rather than continuous seed scheduling.

To fill this gap, we propose a novel corpus reduction technique that can yield a diverse seed corpus subset with minimal runtime overhead. Importantly, the corpus subset can cover all unique program features discovered by the complete corpus. Our key insight is to introduce randomness into corpus reduction so as to enjoy two benefits of a randomized algorithm: randomized output and low runtime overhead. 

Formally, we 
formulate corpus reduction as a \textit{set cover problem}, a classic problem in the combinatorial optimization domain \cite{caprara2000algorithms, karp2009reducibility,chvatal1979greedy}. 
Each seed in the corpus represents a set of program features, while all the program features are represented by the entire seed corpus. The goal of corpus reduction is to find a minimal corpus subset that can cover all the program features.   
Meanwhile, the corpus subset should satisfy three constraints: 1) diverse seed selection; 2) low subset ratio; and 3) low runtime overhead. To solve this problem, we propose a novel randomized set cover algorithm that ·computes a randomized corpus subset (i.e., a different corpus subset at each scheduling round).

We present \sys, a randomized corpus reduction technique that can reduce corpus size and yield diverse seed selection simultaneously. Unlike prior corpus reduction technique (e.g., \texttt{cull\_queue}, AFL-Cmin, MinSet) that produce a deterministic corpus subset, \sys produces a randomized corpus subset in each seed scheduling round, ensuring diverse seed selections. The randomized strategy in \sys greatly reduces runtime overhead in the high-frequency corpus reduction task, compared to traditional set cover algorithms. Moreover, we use compact program features --- frontier node~\cite{she2024fox} rather than edge coverage features to further reduce the size of corpus. Lastly, we build a novel seed scheduling using \sys. Our novel seed scheduling selects a seed within the randomized corpus subset instead of the entire large corpus. By significantly reducing the search space of the seed corpus and encouraging the selection of diverse seeds, our approach effectively mitigates the seed explosion issue and thus improves the fuzzing performance.     

We implement our novel seed scheduling technique on three popular fuzzing frameworks: AFL++~\cite{fioraldi2020afl++}, LibAFL~\cite{fioraldi2022libafl}, and Centipede~\cite{centipede}, demonstrating its generality and practicality. We evaluate its effectiveness using three comprehensive benchmarks: standalone programs, Google's FuzzBench, and the Magma benchmark suite. Our experimental results show that \sys can produce more diverse seed selections compared to the AFL++ and MinSet baseline fuzzers. 
\sys can achieve an average subset ratio of \standsubsetratio and \fuzzbenchsubsetratio after corpus reduction in terms of standalone programs and FuzzBench programs.
In terms of code coverage, \sys achieves \maxstandalonecov higher code coverage in standalone programs compared with AFL++ and \maxaflppfuzzbenchcov, \maxlibaflfuzzbenchcov, and \maxcentipedefuzzbenchcov improvement on FuzzBench targets compared to the three baseline fuzzers (AFL++, Centipede, and LibAFL), respectively. Furthermore, \sys triggers up to \maxmagma more ground-truth bugs compared to the state-of-the-art fuzzer in Magma while introducing minimal runtime overhead (3.93\% in AFL++ on standalone programs and \aflppfuzzbenchover, 9.20\% and 0.35\% in three baseline fuzzers on the FuzzBench dataset).

The main contributions of this paper are:
\begin{itemize}
    \item We investigate the seed explosion problem and corpus reduction technique in fuzzing. 
    \item We formulate corpus reduction in fuzzing seed scheduling as a set cover problem and propose a randomized set cover algorithm to solve this problem.
    \item We design a novel fuzzing seed scheduling that can ~\emph{effectively mitigate} the seed explosion with our randomized corpus reduction. 
    \item We implement our approach in three popular fuzzers and demonstrate its effectiveness through extensive evaluation.  
    \item Our implementation is open-sourced at \url{https://github.com/Crepuscule-v/RandSet}.
\end{itemize}

\section{Background}
In this section, we first explain the main difference between seed scheduling and another similar task in fuzzing: initial seed selection. We then provide a brief introduction to the randomized algorithm and the set cover problem. 
\subsection{Initial Seed Selection vs. Seed Scheduling}
Initial seed selection refers to choosing an optimal subset of initial inputs to form the seed corpus that bootstraps the fuzzing campaign. Mutation-based greybox fuzzing relies on non-crashing seed inputs to initiate bug-finding, and coverage-guided fuzzing efficiency highly depends on initial seed corpus quality \cite{seed_select, rebert2014optimizing, lyu2018smartseed}. Seed scheduling represents the continuous process of selecting seeds from the existing pool for subsequent fuzzing iterations throughout the campaign. This component serves as a critical element of coverage-guided fuzzing, assigning different weights to seed test cases during selection and significantly impacting fuzzing efficiency \cite{She2022EffectiveSS, huang2023balance, wang2021reinforcement, xu2024graphuzz}.

Initial Seed selection and seed scheduling differ in their operational frequency and corresponding runtime cost requirements. While initial seed selection typically occurs once at the campaign's outset to establish the initial corpus, seed scheduling operates continuously throughout the fuzzing process, selecting seeds for each iteration. This difference in task frequency creates contrasting computational requirements---seed scheduling demands much lower runtime cost per operation due to frequent invocation. Consequently, while initial seed selection can afford more expensive algorithms given infrequent execution, seed scheduling requires efficient algorithms with minimal runtime overhead to maintain high fuzzing throughput.

\subsection{Randomized Algorithm}
A randomized algorithm is an algorithm that employs randomness as part of its logic or procedure. Unlike deterministic algorithms, which consistently produce the same output for the same input, randomized algorithms incorporate randomness into their execution flow, resulting in different outputs for the same input~\cite{motwani1996randomized}. Another major benefit of randomized algorithms is runtime efficiency. They typically lead to better expected time complexity or space complexity compared to their deterministic counterparts.

A classic example is \texttt{randomized quicksort}. 
In the worst case, \texttt{deterministic} \texttt{quicksort} can take $O(n^2)$ time when pivot selection consistently results in unbalanced partitions. However, when pivots are selected uniformly at random from the input array, the algorithm achieves $O(n\log n)$ expected running time for any input sequence, where the expectation is taken over the random choices made by the algorithm \cite{cormen2022introduction}.
This same $O(n\log n)$ bound also holds when analyzing the average-case performance over all $n!$ permutations of $n$ distinct elements with equal probability. Thus, randomization effectively eliminates the algorithm's dependence on adversarial input orderings.

\subsection{Set Cover Problem}
The set cover problem is a classic problem in computer science~\cite{caprara2000algorithms, karp2009reducibility,chvatal1979greedy}. The goal of the set cover problem is to find a minimum number of subsets to include every unique element of all subsets. Formally, given a universal set of elements \replaced{$U = \{e_1, e_2, ..., e_n\}$}{$\mathcal{F} = \{f_1, f_2, ..., f_m\}$} and a collection of subsets \replaced{$S = \{S_1, S_2, ..., S_m\}$}{$\mathcal{A} = \{A_1, A_2, ..., A_n\}$} where each \replaced{$S_i \subseteq U$}{$A_i \subseteq \mathcal{F}$}, the goal is to find a minimum number of subsets $C$ such that the union of all subsets in \replaced{$C$}{$\mathcal{C}$} equals \replaced{$U$}{$\mathcal{F}$}:
\begin{equation}
\replaced{\bigcup_{S_i \in C} S_i = U}{\bigcup_{A_i \in \mathcal{C}} A_i = \mathcal{F}}
\end{equation}

The set cover problem is known to be NP-hard~\cite{karp2009reducibility}. However, there exists a greedy algorithm that achieves a logarithmic approximation ratio of \replaced{$H(n)$}{$H(m)$}, where \replaced{$H(n)$ is the $n$-th harmonic number and $n$ is the size of the universal set}{$H(m)$ is the $m$-th harmonic number and $m$ is the size of the universal set}~\cite{chvatal1979greedy}. At each time step, this greedy algorithm selects the set that covers the maximum number of remaining uncovered elements:
\begin{equation}
\replaced{S_t = \mathop{\argmax}_{S_i \in S} |S_i \cap U_t|}{A_t = \mathop{\argmax}_{A_i \in \mathcal{A}} |A_i \cap U_t|}
\end{equation}
where $U_t$ represents the set of elements that remain uncovered at time step $t$. The approximation ratio of \replaced{$H(n)$}{$H(m)$} is proven to be optimal under standard complexity assumptions, as it has been shown that achieving an approximation ratio of \replaced{$(1-\epsilon)\ln n$}{$(1-\epsilon)\ln m$} for any $\epsilon > 0$ is NP-hard~\cite{feige1998threshold}. This theoretical foundation makes the greedy approach particularly attractive for practical applications where optimal solutions are computationally infeasible.

\section{Corpus Reduction}
\label{sec:challenge}
The seed explosion problem fundamentally stems from the monotonic growth of the seed corpus during long-running fuzzing campaigns. As fuzzers continuously generate new seeds through mutation and discover new code coverage, the corpus size can grow from hundreds to tens of thousands of seeds over extended periods. This exponential growth creates a scalability bottleneck where fuzzers must maintain an increasingly large pool of candidates at each scheduling round. Consequently, promising seeds that could lead to new discoveries become buried in the ever-expanding corpus, significantly reducing fuzzing efficiency.

Current seed scheduling approaches \cite{ecofuzz,lemieux2018fairfuzz, aflfast, slowfuzz, wang2021reinforcement, She2022EffectiveSS, Wang2020NotAC} primarily focus on developing sophisticated prioritization strategies to rank seeds based on various metrics such as coverage feedback, energy assignment, or reinforcement learning. However, these prioritization-based solutions fail to address the root cause of seed explosion, as they still operate on the entire corpus. The large seed corpus leads to a typical long-tail effect, where it is challenging to select a promising seed.

A better approach is to apply corpus reduction techniques that actively reduce the size of the seed corpus while preserving its essential characteristics. 
To successfully apply corpus reduction for seed scheduling, several key challenges must be addressed. We first identify these challenges and then analyze why existing corpus reduction techniques fail to meet these requirements.

\subsection{Challenges in Practical Application}
\noindent\textbf{Seed diversity} is particularly crucial for seed scheduling in fuzzing. Diversity among seeds ensures that the fuzzer explores a broader range of program behaviors, thereby increasing the likelihood of discovering new paths and vulnerabilities. However, corpus reduction inevitably reduces the original seed corpus to a much smaller subset which leaves most seeds discarded. Therefore, to preserve seed diversity in the fuzzing process, it is essential to design a non-deterministic approach when selecting the subset, so that the selected corpus subset for fuzzing can vary across runs. Otherwise, similar seed corpora will consistently yield similar subsets, causing the seed scheduler to be restricted to a small, fixed set of seeds. This greatly wastes computing resources, especially evident when fuzzing gets stuck.

\noindent\textbf{Subset Ratio} is another core metric for evaluating the corpus reduction algorithms. The motivation for corpus reduction is to address the seed explosion problem during fuzzing. Thus, if the subset ratio is inadequate, the fuzzer is still forced to perform seed scheduling on a redundant corpus, which undermines the original intent of corpus reduction. Therefore, a low subset ratio is critical to the effectiveness of the algorithm in real-world fuzzing campaigns.

\noindent\textbf{Runtime overhead} is a decisive factor in determining whether a technique can be practically applied in high-frequency seed scheduling scenarios. For initial seed selection related work, the operation is typically performed only once before fuzzing starts, so runtime overhead is not a significant concern. In contrast, our goal is to alleviate the seed explosion problem during the fuzzing process by maintaining a small, high-quality subset for seed scheduling. Therefore, low runtime overhead becomes an essential consideration in the design of such techniques.

\subsection{Limitations of Existing Techniques}
\label{limit_of_existing_tech}

We conducted a comprehensive review of existing corpus reduction techniques from both academia and industry. In general, these techniques can be divided into two categories: offline corpus reduction approaches (e.g. AFL-Cmin \cite{aflcmin}, MinSet\cite{rebert2014optimizing, peachminset}), which are primarily applied in the initial seed selection stage before fuzzing starts, and online corpus reduction approaches (e.g. \texttt{cull\_queue} \cite{aflcullqueue} in AFL/AFL++), which are used during the seed scheduling stage. However, none of the existing techniques are practical for corpus reduction during the seed scheduling stage. The performance of these techniques with respect to the three challenges is summarized in \autoref{tab:challenge}. It is evident that none of them can satisfy all three requirements simultaneously. In the following, we analyze the limitations of each tool separately.

\textbf{Limitation of MinSet.} 
The limitations of MinSet can be attributed to both its deterministic design and its reliance on less expressive features for subset selection. First, MinSet typically employs a greedy set cover algorithm for subset selection, yielding deterministic solutions. As a result, the selected subset is relatively fixed, which restricts the diversity of seeds available for subsequent scheduling. Ultimately, this reduces overall seed diversity during fuzzing. Additionally, this algorithm requires repeatedly traversing the entire seed corpus to select the best seed (i.e., the one that covers the most remaining uncovered basic blocks), which becomes extremely time-consuming as the corpus scales up. Moreover, the subset selection process relies on features with large feature size, such as basic blocks or edges. While this approach can reduce the corpus size to some extent, the limited expressiveness of these features often results in an insufficient subset ratio.

\textbf{Limitation of AFL-Cmin.}
Several shortcomings noted for MinSet are also shared by AFL-cmin, such as the lack of diversity caused by deterministic processing, the limited subset ratio associated with large feature spaces, and the fact that both are designed as offline, one-time corpus reduction tools, with such task scenarios inherently neglecting considerations of runtime efficiency. Beyond these shared limitations, AFL-cmin further suffers from significant runtime overhead due to expensive disk I/O operations, which hinders its applicability in high-frequency seed scheduling scenarios.

\begin{table}[t!]
\centering
    \caption{\textbf{Comparison of existing corpus reduction techniques across three key challenges.\emptycirc\ indicates the challenge is not addressed, \halfcirc\ indicates the challenge is partially addressed, and \fullcirc\ indicates the challenge is well addressed.}}   
    \label{tab:challenge}
    \renewcommand{\arraystretch}{1.5}
    \scalebox{0.9}{ 
\begin{tabular}{l|ccc}
\toprule
\textbf{Challenge}                     & \textbf{AFL-Cmin} & \textbf{MinSet} & \textbf{cull\_queue} \\ \hline
\textbf{Seed Diversity} & \emptycirc       & \emptycirc         & \emptycirc          \\ \hline
\textbf{Subset Ratio} & \halfcirc       & \halfcirc          & \halfcirc          \\ \hline
\textbf{Runtime Overhead}                & \emptycirc       & \emptycirc         & \fullcirc    \\
 \bottomrule
\end{tabular}
}
\end{table}

\textbf{Limitation of cull\_queue.}
\label{limit_of_cull_queue} 
Considering that \texttt{cull\_queue} is the technique most closely aligned with our task scenario, we will elaborate on why it is impractical in practice based on our source code analysis and empirical study results. The \texttt{cull\_queue} can be seen as a two-stage process. In the first stage, it selects a corpus subset that collectively covers all the edges covered so far and marks these seeds as \textit{favored}. Concretely, the algorithm iterates over all edges in order. For each edge not yet covered by the current subset, it selects the corresponding \textit{top\_rated} seed (i.e., the seed with the smallest execution time and file length) and adds it to the subset, continuing until all edges are covered. In the second stage, the first \textit{favored} seed that has not yet been fuzzed is marked as \textit{smallest\_favored} and chosen as the final candidate for mutation. Therefore, the entire algorithm is deterministic: given the same seed corpus, it will always yield the same \textit{smallest\_favored} seed.

\begin{table}[!ht]
\centering
\footnotesize
\caption{\textbf{The average number of seeds selected via the cull\_queue (CQ) and \textit{weighted random} (WR) strategies over all scheduling rounds during a 1-hour AFL++ fuzzing campaign (10-run average, across both FuzzBench and standalone targets).}}
\label{tab:cull_vs_random}
\renewcommand{\arraystretch}{1.1}
\setlength{\tabcolsep}{4pt}
\begin{tabular}{lrr || lrr || lrr }
\toprule
\textbf{Targets} & \textbf{CQ} & \textbf{WR} & \textbf{Targets} & \textbf{CQ} & \textbf{WR}  & \textbf{Targets} & \textbf{CQ} & \textbf{WR}\\
\midrule
\multicolumn{9}{c}{\textbf{FuzzBench Targets}} \\
\midrule
bloaty      & 16.6  & 3327.4   & openh264   & 91.6  & 340.4   & libxslt  & 159.8 & 28904.2 \\
curl            & 88.8  & 67496.2  & openssl        & 24.6  & 73592.0 & libxml     & 179.2 & 21687.0 \\
harfbuzz        & 479.6 & 17167.2  & re2       & 100.6 & 18222.6 & zlib           & 6.6   & 2362.0 \\
jsoncpp         & 27.0  & 59271.2  & sqlite3 & 29.8  & 16781.8 & libpng     & 15.4  & 68039.8 \\ \cline{7-9}
lcms            & 5.2   & 27049.2  & stbi           & 43.2  & 2757.6 & \multirow{3}{*}{\textbf{Mean Ratio}} & \multirow{3}{*}{\textbf{1.61\%}} & \multirow{3}{*}{\textbf{98.39\%}} \\
libjpeg         & 63.2  & 25411.4  & vorbis         & 21.8  & 49471.8 & & &\\
libpcap         & 1.0   & 242978.4 & woff2          & 28.4  & 74966.2 & & &\\
\midrule
\multicolumn{9}{c}{\textbf{Standalone Targets}} \\
\midrule
bsdtar      & 30.3  & 921.7   & exiv2    & 5.0   & 1460.3 & xmllint & 61.3 & 915.3 \\
ffmpeg      & 61.0  & 128.3   & jhead    & 6.7   & 4198.0 & tcpdump  & 124.0 & 333.3 \\
jq          & 30.3  & 516.0   & mp3gain  & 23.3  & 1552.0 & strip-new   & 44.3  & 973.7\\ \cline{7-9}
mp42aac     & 17.3  & 905.7   & nm-new   & 33.0  & 1000.0 & \multirow{3}{*}{\textbf{Mean Ratio}} & \multirow{3}{*}{\textbf{10.16\%}} & \multirow{3}{*}{\textbf{89.84\%}} \\
objdump     & 78.7  & 341.0   & pdftotext & 55.0 & 435.0  & & &  \\
readelf     & 99.3  & 212.7 & size     & 42.0  & 830.0 & & & \\
\bottomrule
\end{tabular}
\end{table}

More importantly, during our analysis of the source code, we also found that due to the design of AFL++, the condition for selecting a seed for mutation via \texttt{cull\_queue} is highly stringent. First, if no better seed (i.e., smaller and faster) is found in a new fuzzing round, the \texttt{cull\_queue} performs an early return. This implies that, as coverage plateaus,  the fuzzer relies almost entirely on the weighted random strategy to sample from the full seed corpus. Second, even when a \textit{smallest\_favored} seed is eventually selected by \texttt{cull\_queue}, it may still be skipped if it has not yet had its \textit{perf\_score} computed. 

Furthermore, our empirical study substantiates our analysis. We conducted extensive experiments on both FuzzBench and standalone targets. \autoref{tab:cull_vs_random} reports the average number of seeds selected through \texttt{cull\_queue} and weighted random strategies across all scheduling rounds in a 1-hour AFL++ fuzzing campaign. Surprisingly, \texttt{cull\_queue} was employed in only 1.61\% of scheduling rounds for FuzzBench targets and 10.16\% for standalone targets. This reveals that the \texttt{cull\_queue} strategy is barely adopted during real-world fuzzing. As a result, in the vast majority of cases, seed scheduling is performed over the entire seed corpus, thus inevitably suffering from seed explosion. Therefore, a novel and practical solution for corpus reduction in the seed scheduling stage is urgently needed.
\section{Corpus Reduction as a Set Cover Problem}
\label{section:formulation}
In this section, we introduce the details of the problem formulation: corpus reduction as a set cover problem in seed scheduling. Given the inevitable redundancy in the seed corpus due to the seed-saving policy of fuzzing, our goal is to find a minimal subset of seeds that preserves all unique features of the entire corpus while satisfying the stringent requirements of dynamic seed scheduling scenarios. We identify the relationship between seeds and their corresponding features, then formally model the corpus reduction as a set cover problem.

We first explain the root cause of seed corpus redundancy during a fuzzing campaign. According to the widely-used seed-saving policy~\cite{fioraldi2020afl++, afl, fioraldi2022libafl, centipede}, a new seed is saved when it triggers previously undiscovered features, which can be defined as dynamic program execution behaviors such as basic blocks, control flow edges, or unique execution counts on these features. As the fuzzing campaign continues, the seed corpus continues to grow. However, in this process, a lot of redundancy arises because each seed can represent \emph{multiple} features. A later saved seed that represents a new feature and multiple old features causes a previously saved seed to be redundant.



The relationship between seeds and features exhibits a many-to-many mapping characteristic. A single seed can trigger multiple features, while multiple seeds can trigger a specific feature. 
To formally describe this relationship, let $S = \{s_1, s_2, \ldots, s_n\}$ denote the set of $n$ seeds in the corpus, and \replaced{$F = \{f_1, f_2, ..., f_m\}$}{$\mathcal{F} = \{f_1, f_2, ..., f_m\}$} represent the set of $m$ features discovered during fuzzing. For each seed $s \in S$, we define its feature set \replaced{$F(s) \subseteq F$}{$\Phi(s) \subseteq \mathcal{F}$}  as the set of features it covers.

\begin{figure}
    \centering
    \includegraphics[width=0.6\linewidth]{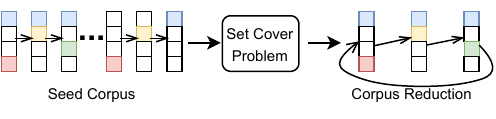}
    \caption{\textbf{An example of our seed corpus reduction. We formulate the corpus reduction as a set cover problem. The seed corpus with redundant seeds (6 seeds) is reduced into a distilled corpus (3 seeds). The distilled corpus preserves all feature coverage (4 colors) with minimal redundancy. Each vertical bar represents a seed and each colored block indicates a unique feature covered by the seeds.}}
    \label{fig:formulation}
\end{figure}

To mitigate the impact of redundancy, we propose distilling the seed corpus into a minimal but expressive subset. As illustrated in \autoref{fig:formulation}, our approach takes the original seed corpus as input, where each seed (represented as a vertical bar) is associated with multiple features. Each unique color block represents one corresponding feature. We formulate the corpus reduction as a set cover problem, which processes the seed-feature mappings to identify the minimal set of seeds that maintain complete feature coverage. The output is a distilled corpus that preserves all the discovered features while eliminating redundant seeds. Formally, our goal is to find a minimal subset $S' \subseteq S$ such that
\begin{equation}
    \replaced{\bigcup_{s \in S'} F(s) = \bigcup_{s \in S} F(s)}{\bigcup_{s \in S'} \Phi(s) = \mathcal{F}}
\end{equation}
\replaced{, meaning that all seed in the subset $S'$ cover all unique features discovered by the original seed corpus $S$, represented as follows.}{, meaning that all seeds in the subset $S'$ collectively cover all unique features in $\mathcal{F}$ discovered by the original seed corpus $S$, represented as follows.}

\section{Methodology}

Building upon the set cover problem formulation in the previous section, we propose RandSet, a novel randomized corpus reduction technique for seed scheduling. RandSet addresses the fundamental challenge of balancing corpus diversity and computational efficiency by introducing randomness into the reduction process. Our key insight is that randomization provides two critical benefits: diverse seed selection through randomized output and minimal runtime overhead suitable for high-frequency seed scheduling tasks.

In this section, we first explain why classic deterministic set cover algorithms are insufficient for seed scheduling. We then present our randomized corpus reduction technique, which consists of two key innovations: a randomized set cover algorithm that employs randomization to break deterministic patterns while reducing computational complexity, and the adoption of frontier nodes as the core feature for corpus reduction, which enhances the subset ratio by focusing on boundary regions between explored and unexplored program areas. Furthermore, we showcase a novel seed scheduling framework that demonstrates the real-world implications of RandSet.

\begin{figure*}[ht]
\centering
\includegraphics[width=0.9\linewidth]{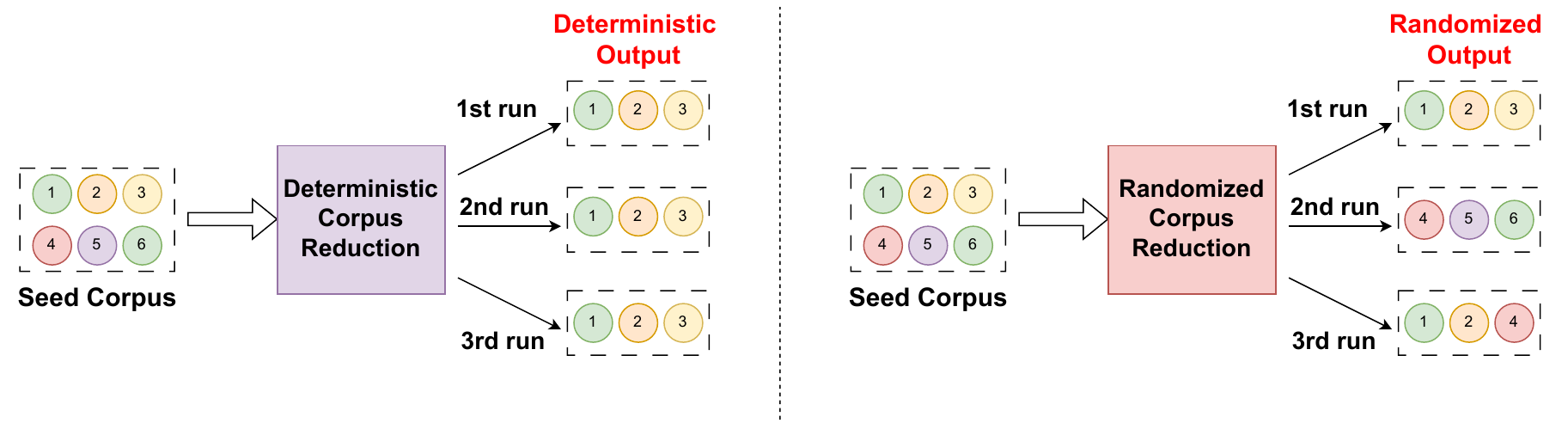}
    \caption{\textbf{Diversity comparison between Deterministic and Randomized Corpus Reduction algorithms.} We demonstrate the selection results from a seed corpus of 6 seeds using different reduction algorithms. Deterministic Corpus Reduction algorithms, i.e., MinSet, AFL-Cmin, and \texttt{cull\_queue} (left) consistently select the same three seeds (1, 2, 3) across multiple runs, while our Randomized Set Cover algorithm (right) produces diverse selections: seeds (1, 2, 3) in the first run, seeds (4, 5, 6) in the second run, and seeds (1, 2, 4) in the third run. Each numbered circle represents a seed.
    }
\label{fig:diversity}
\end{figure*}

\subsection{Why Not Existing Set Cover Algorithms?}
\label{sub:determin}
In this subsection, we analyze the fundamental limitations of employing existing set cover algorithms, including both classic deterministic and LP-based rounding approaches.

\textbf{Classic Deterministic Algorithm}
The classic deterministic set cover algorithm is essentially a greedy algorithm. First, the deterministic nature leads to a seed diversity problem. During the fuzzing process, there exist periods where no new features are discovered, resulting in no new seeds being added to the seed corpus. In such scenarios, as shown in the left part of Figure \ref{fig:diversity}, if a deterministic set cover algorithm is employed, it will consistently produce identical results (seeds 1, 2, 3) across multiple iterations. This lack of diversity limits the exploration potential of the fuzzer and may cause it to get stuck in local optima.

Additionally, there is a substantial computational overhead problem. The greedy algorithm requires multiple iterations to find the solution. In each iteration, it must scan all \replaced{$m$}{$n$} candidate sets and examine each feature-set relationship to determine which set covers the maximum number of uncovered features. This results in a time complexity of \replaced{$O(mn^2)$}{$O(n^2m)$}, \replaced{$m$ is the number of candidate sets and $n$ is the total number of features}{$n$ is the number of candidate sets and $m$ is the total number of features}. This computational overhead becomes particularly problematic in fuzzing scenarios where corpus reduction needs to be performed frequently and efficiently.

\textbf{LP-based rounding Algorithm}
To address the limitations of the classic greedy approach, one might consider LP-based (Linear Programming) rounding set cover algorithms\cite{vazirani2001approximation}, which include both standard and randomized variants. While randomized rounding could alleviate the aforementioned diversity issue, all LP-based approaches introduce prohibitive runtime overhead by requiring an external LP solver to compute relaxations. We provide a detailed empirical comparison evaluating both standard and randomized LP-based rounding algorithms later in \autoref{sub:LP_based_rounding}.

\subsection{Randomized Set Cover Algorithm}
In this subsection, we present our proposed Randomized Set Cover Algorithm for seed scheduling and demonstrate how it addresses the diversity and runtime overhead limitations. We first introduce the algorithm's design, then analyze how its randomized approach effectively tackles both the lack of diversity and the computational overhead in deterministic set cover algorithm.

Algorithm \autoref{alg:dynamic} presents our Randomized Set Cover Algorithm for seed scheduling. The algorithm takes three inputs: the global feature set \texttt{FEATURES} containing all discovered features, the seed queue \texttt{SEED}, and \replaced{the feature sets $S$ where $S_i$ represents the features covered by $seed_i$}{the feature mapping $\Phi(\cdot)$ where $\Phi(seed_i)$ represents the features covered by $seed_i$}. The goal is to return a reduced set of seeds that maintains coverage while introducing diversity.




\begin{algorithm}[h]
\small
\caption{Randomized Set Cover Algorithm for Efficient and Diverse Corpus Reduction} 
\label{alg:dynamic} 
\begin{algorithmic}[1]
\Require \replaced{Global feature set $FEATURES$, Seed queue $SEED$, Feature set $S = \{S_1, S_2, ..., S_n\}$, where $S_i$ denotes the set of features covered by $seed_i$}{Global feature set $\mathcal{F}$ (\texttt{FEATURES}), Seed queue $SEED$, and feature mapping $\Phi(\cdot)$, where $\Phi(seed_i) \subseteq \mathcal{F}$ denotes the set of features covered by $seed_i$}
\Ensure Distilled seed corpus \replaced{$R$}{$\mathcal{R}$}
\State \replaced{$C \gets FEATURES$}{$U \gets \mathcal{F}$} \Comment{\textcolor{purple}{\replaced{Set of features yet to be covered}{Set of uncovered features}}}
\State \replaced{$R \gets \emptyset$}{$\mathcal{R} \gets \emptyset$}
\State $SEED' \gets shuffle(SEED)$ \Comment{\textcolor{purple}{Introduce randomness}}

\For{each seed $seed_i \in SEED'$}
    \If{\replaced{$S_i \cap C \neq \emptyset$}{$\Phi(seed_i) \cap U \neq \emptyset$}} \Comment{\textcolor{purple}{\replaced{$S_i$ covers new feature}{$\Phi(seed_i)$ covers new feature}}}
        \State \replaced{$R \gets R \cup \{seed_i\}$}{$\mathcal{R} \gets \mathcal{R} \cup \{seed_i\}$} \Comment{\textcolor{purple}{\replaced{Add $S_i$ to set cover}{Add $seed_i$ to the cover}}}
        \State \replaced{$C \gets C \setminus S_i$}{$U \gets U \setminus \Phi(seed_i)$} \Comment{\textcolor{purple}{\replaced{Update set of features}{Update set of uncovered features}}}
    \EndIf
    \If{\replaced{$C = \emptyset$}{$U = \emptyset$}} 
        \State \textbf{break} 
    \EndIf
\EndFor

\State \Return \replaced{$R$}{$\mathcal{R}$}
\end{algorithmic}
\end{algorithm}

The algorithm operates as follows. First, it initializes \replaced{$C$}{$U$} as using \replaced{\texttt{FEATURES}}{$\mathcal{F}$ (\texttt{FEATURES})} which represents the set of features that still need to be covered (line 1). It then creates an empty set \replaced{$R$}{$\mathcal{R}$} to store the selected seeds (line 2). A key innovation is introduced in line 3, where we shuffle the seed queue to introduce randomness into the selection process. This shuffling ensures that seeds have the same probability to be considered first, breaking the deterministic nature of traditional greedy algorithms of the set cover problem.

The main selection process (lines 4--10) iterates through the shuffled seed queue. For each seed, if it covers any yet-uncovered features (line 5), it is added to the result set \replaced{$R$}{$\mathcal{R}$} (line 6), and its covered features are removed from \replaced{$C$}{$U$} (line 7). This process continues until either all seeds have been considered or all features have been covered (lines 8--10). Finally, the algorithm returns the distilled seed corpus \replaced{$R$}{$\mathcal{R}$} that contains the selected seeds.

Our Randomized Set Cover algorithm directly tackles the first challenge identified in Section \ref{sec:challenge} --- seed diversity. As shown in the right part of Figure \ref{fig:diversity}, our proposed Randomized Set Cover algorithm generates different reduction subsets even when operating on the same seed corpus, producing varied selections such as (1, 2, 3), (4, 5, 6), and (1, 2, 4) across different runs. Additionally, there is a relatively rare scenario where the corpus evolves with new seeds and features. In such cases, the deterministic algorithm often continues to select overlapping subsets, leading to repetitive selections. In contrast, our randomized approach maintains diversity by selecting different combinations across scheduling rounds. This randomization ultimately enhances the diversity of seeds selected for execution, leading to more comprehensive exploration of the program's execution space.

Our algorithm also addresses the second challenge --- the computational overhead. Compared to the deterministic approach discussed in subsection~\ref{sub:determin}, our Randomized Set Cover algorithm employs a much simpler approach: it first shuffles the \replaced{$m$}{$n$} sets using Fisher-Yates algorithm in \replaced{$O(m)$}{$O(n)$} time, then performs a single pass through all sets, checking each feature-set relationship exactly once using hash table operations. This results in a reduced time complexity of \replaced{$O(mn)$}{$O(nm)$}, \replaced{$m$ is the number of candidate sets and $n$ is the total number of features}{$n$ is the number of candidate sets and $m$ is the total number of features}, eliminating the multiplicative factor of $n$ present in the deterministic approach. Since our algorithm eliminates this multiplicative factor, it achieves substantial performance improvements, especially when dealing with large seed corpora where $n$ can be significant. The empirical comparison of runtime overhead between our approach and the greedy algorithm is presented in Section~\ref{sec:evaluation}.

\subsection{Frontier Node Feature}
In this subsection, we present our approach for selecting an effective feature to guide corpus reduction. Building upon our Randomized Set Cover algorithm that addresses diversity and computational overhead, we tackle the third challenge of achieving lower subset ratio by adopting frontier nodes as the key feature for our seed scheduling algorithm. This choice leverages their unique properties as boundary indicators between explored and unexplored program regions, enabling more aggressive corpus reduction while maintaining exploration effectiveness.

Previous works ~\cite{She2022EffectiveSS, she2024fox, centipede} introduced the concept of frontier nodes in fuzzing. A frontier node is defined as a visited control flow node that has at least one unvisited child node in the program's Control Flow Graph (CFG). These frontier nodes collectively form the boundary between explored and unexplored regions of the program's control flow, effectively identifying promising areas for further exploration during coverage-guided fuzzing campaigns.

Our choice of frontier nodes as the feature for seed scheduling offers several advantages over edge-based approaches, particularly in achieving lower subset ratio. Let $SS_1$ denote the subset of seeds scheduled using edges as features, and $SS_2$ denote the subset scheduled using frontier nodes as features. For any seed $s_i \in SS_1$, let $E_i$ represent the set of edges covered by $s_i$, and $F_i$ represent the set of frontier nodes covered by $s_i$. \replaced{Since frontier nodes are defined as visited nodes with unvisited children, we have $F_i \subseteq E_i$ for all seeds.}{Since frontier nodes are defined as visited CFG nodes with unvisited children, we have $F_i \subseteq V_i$ for all seeds, where $V_i$ denotes the set of CFG nodes visited by $s_i$ (which can be derived from $E_i$).} However, not all edges in $E_i$ necessarily contribute to frontier node coverage, meaning some edges may cover already-explored regions without advancing the exploration boundary.

\replaced{Formally, there exists a subset $SS_2 \subseteq SS_1$ such that $\bigcup_{s_i \in SS_2} F_i = \bigcup_{s_j \in SS_1} F_j$, where $|SS_2| \leq |SS_1|$.}{Formally, let $SS_1^\star$ be a minimum-size seed subset that covers all edges, and let $SS_2^\star$ be a minimum-size seed subset that covers all frontier nodes. Since covering all edges implies visiting all frontier nodes, any feasible edge cover is also feasible for frontier-node coverage, and thus $|SS_2^\star| \leq |SS_1^\star|$.} This relationship demonstrates that frontier node-based scheduling can achieve equivalent exploration coverage with a potentially smaller number of seeds, directly improving the subset ratio of corpus reduction.

Additionally, frontier nodes offer computational advantages that further enhance reduction efficiency due to their smaller quantity compared to the total number of edges in a program's control flow graph. Since the set cover algorithm's complexity depends on the size of the universe being covered, using frontier nodes as features reduces the computational overhead during the scheduling process. This efficiency gain becomes particularly significant for large programs with extensive control flow graphs, where $|FRONTIER| \ll |EDGES|$, leading to faster convergence of the set cover algorithm. Specifically, \replaced{assuming the number of frontier nodes is $k$ and the number of edges is $s$, the complexity reduces from $O(ns)$ to $O(nk)$}{assuming the number of frontier nodes is $k$ and the number of edges is $L$, the complexity reduces from $O(nL)$ to $O(nk)$}, where $n$ is the number of candidate sets, resulting in substantial performance improvements.

\subsection{Novel Seed Scheduling with Randomized Corpus Reduction}

In this subsection, we build a novel seed scheduling using RandSet, our randomized corpus reduction technique, that can effectively mitigate the seed explosion. Algorithm~\ref{alg:scheduling} presents our approach, which integrates RandSet into the fuzzing loop to maintain both diversity and efficiency. During each fuzzing iteration, we first apply our RandSet algorithm to obtain a reduced subset of seeds that preserves feature coverage while significantly reducing the corpus size. This subset represents a diverse yet manageable collection of seeds that maintains the essential characteristics of the full corpus.

\begin{algorithm}[h]
\small
\caption{Novel Seed Scheduling with Randomized Corpus Reduction}
\label{alg:scheduling}
\begin{algorithmic}[1]
\Require Seed corpus $CORPUS$
\While{fuzzing not terminated}
\State $SUBSET \gets RandSet(CORPUS)$ \Comment{\textcolor{purple}{Reduced subset}}
\State $seed \gets SelectSeed(SUBSET)$ \Comment{\textcolor{purple}{Select from subset}}
\State $MutateAndExecute(seed)$
\added{\Comment{\textcolor{purple}{Generate multiple mutants and execute}}}
\EndWhile
\end{algorithmic}
\end{algorithm}

For the SelectSeed function, similarly to previous prioritization techniques~\cite{aflfast,lemieux2018fairfuzz}, we choose the newest seed within the list, as recent seeds often exhibit higher potential for discovering new program behaviors. The selected seed then undergoes mutation and execution, with coverage information updated to reflect any newly discovered program states. This approach ensures that our fuzzing process benefits from both the computational efficiency of corpus reduction and the exploration effectiveness of prioritized seed selection.

\section{Implementation}
We implement our approach in \sys, a corpus reduction system that can be integrated into different fuzzing frameworks. \sys consists of two main components: (1) a frontier node identification module that constructs the \replaced{inter-procedural}{intra-procedural} Control Flow Graph (CFG) and identifies frontier nodes, and (2) a seed selection module that implements the set cover algorithm. 
We have integrated \sys into \aflpp, \libafl and \centipede to demonstrate its broad applicability.

\subsection{Control Flow Graph Construction}
\label{cfg_construction_description}
To classify CFG nodes as visited, we reuse a fuzzer's edge coverage information to identify visited basic blocks. However, the method of obtaining the \replaced{inter-procedural CFG}{intra-procedural CFG} differs between AFL++ and LibAFL integrations.

\textbf{AFL++ Integration.} For AFL++, we modified its instrumentation pass to directly extract basic block level \replaced{inter-procedural CFG}{intra-procedural CFG} information during compilation. Specifically, we extended AFL++'s LLVM instrumentation mode to collect control flow information between basic blocks. This approach allows us to construct the \replaced{inter-procedural CFG}{intra-procedural CFG} without requiring additional external tools.

\textbf{LibAFL Integration.} \replaced{For LibAFL, we utilize LLVM's infrastructure to construct the inter-procedural CFG. We first compile the target program with gllvm and use LLVM's opt tool to extract each function's intra-procedural CFG. We then implement a LLVM pass to process these individual CFGs and merge them based on function call relationships identified in the LLVM IR. This process runs on LLVM version llvm-15 and produces a complete inter-procedural CFG that captures both intra-function control flow and inter-function call relationships.}{For LibAFL, we leverage LLVM's infrastructure to construct the intra-procedural CFG for each function. We first compile the target program with gllvm, and then utilize a modified LLVM pass to traverse the LLVM IR and extract basic block-level control flow information. This process, executed on LLVM-15, results in a collection of independent CFGs representing the internal control flow of each function.}

\textbf{Centipede Integration.} Centipede assigns a global index to each basic block based on the \texttt{trace-pc-guard} instrumentation. In addition, LLVM 16 introduces a new feature, enabled via the \texttt{-fsanitize-coverage=control-flow} flag, which emits control flow information into the \texttt{.sancov\_cfs} section. Centipede leverages this control flow coverage data to enable advanced fuzzing features such as coverage frontier analysis. In our evaluation on centipede, all targets are compiled and executed using LLVM 18.

\subsection{Algorithm Implementation}
To realize the dynamic set cover process, an index list consists of random numbers is created all at once instead of calling the ``random()'' function to create a random seed index many times. Then, we pick up the seed by traversing the list and calculate the intersection number between its  covered frontier nodes and global frontier nodes bitmap. We further remove its covered frontier nodes from global frontier node bitmap until the total of intersection number is equal to the global number of frontier nodes, which means all frontier nodes are covered by the seeds in reduced set.

\section{Evaluation}
\label{sec:evaluation}

In this section, we aim to answer the following research questions:

\begin{itemize}
    \item \textbf{RQ1:} Can \sys improve the diversity of the seeds selected for fuzzing compared to the baseline methods? (Diversity)
    \item \textbf{RQ2:} How effective is \sys in reducing the size of the seed corpus? (Reduction)
    \item \textbf{RQ3:} How well does \sys perform in code coverage compared with state-of-the-art fuzzers? (Coverage)
    \item \textbf{RQ4:} How many unique bugs can \sys discover compared to baseline fuzzers? (Bug Discovery)
    \item \textbf{RQ5:} \replaced{How does the runtime overhead of corpus reduction in \sys compare to that of other baseline methods? (Runtime Overhead)}{What is the performance overhead of \sys, considering both the offline CFG construction and the runtime subset generation? (Performance Overhead)}
    \item \textbf{RQ6:} Can \sys generalize to other fuzzing engines, such as \libafl and \centipede ? (Generalizability)
    \item \textbf{RQ7:} How does each component of \sys contribute to its overall effectiveness? (Ablation Study)
    \item \added{\textbf{RQ8:} Can other existing randomized approximation algorithms for Set Cover serve as effective alternatives to RandSet’s approach? (Alternative Set Cover Algorithms) }
\end{itemize}

\textbf{Baselines.}
We evaluate \sys through its integration with the state-of-the-art fuzzing framework AFL++ (version 4.10c) \cite{fioraldi2020afl++}.
AFL++ represents the latest iteration of the widely-recognized AFL family of fuzzers \cite{asprone2022comparing}, consistently demonstrating superior performance and topping the January 2025 FuzzBench report \cite{aflpp_fuzzbench_report}. 
We adopt \sys into AFL++ and use vanilla AFL++ as a primary baseline for comparison. Additionally, we extend our comparison to three other baselines to more thoroughly evaluate our method against prior corpus reduction techniques. Including: 
\begin{enumerate}
    \item \textbf{\aflppcmin}: AFL++ integrates periodic corpus minimization by invoking afl-cmin via \texttt{execve}. Since determining the optimal interval is a challenging tuning task and outside the scope of our study, we simply adopt a fixed 5-minute interval in our experimental setting.
    \item \textbf{\aflppcq}: AFL++ uses only the \texttt{cull\_queue} mechanism for corpus reduction and seed selection, with the built-in weighted random selection strategy disabled.
    \item \textbf{\aflppminset}: AFL++ uses a greedy set cover algorithm to select a minimal subset of seeds for seed scheduling.
\end{enumerate}
 We specifically choose these three baselines because they represent the current mainstream corpus reduction approaches and are most comparable to our method. \added{Furthermore, all experiments were conducted using the default AFL++ configuration, without enabling \texttt{CmpLog}.}

\textbf{Benchmark Programs}
For our evaluation, we use the FuzzBench dataset~\cite{metzman2021fuzzbench}, which has become a standard benchmark in fuzzing research~\cite{gorz2024sbft, asprone2022comparing, bohme2022reliability}. We successfully compiled \numfuzzbenchaflpp targets from the FuzzBench dataset, while excluding a few targets due to compilation failures or dependency issues.  
To provide a more comprehensive evaluation, we additionally selected 15 real-world standalone programs from recent work~\cite{she2024fox,zhang2023shapfuzz,li2021UniBench}, spanning diverse domains such as multimedia processing, document parsing, compression tools and so on. These targets feature larger code sizes and more complex control flows, and were used for testing with \aflpp. 
The complete list of evaluation targets is presented in \autoref{tab:programs}.


\begin{table}[]
\centering
\footnotesize
\caption{\textbf{Studied programs in our evaluation. Programs marked with * failed to compile with \libafl, while those marked with $\ddagger$ failed to compile with \centipede.}}
\label{tab:programs}
\renewcommand{\arraystretch}{1.1}
\begin{tabular}{ll|ll|ll|ll}
\toprule
\textbf{Targets}  & \textbf{Version} & \textbf{Targets}  & \textbf{Version} & \textbf{Targets}  & \textbf{Version} & \textbf{Targets}  & \textbf{Version} \\
\midrule
\multicolumn{2}{c|}{\textbf{FuzzBench Targets}} & re2           & b025c6a & woff2         & 8109a2c & sqlite3        & c78cbf2\\ \cline{1-2}
bloaty        & 52948c1 & curl*$\ddagger$ & a20f74a & stbi           & 5736b15 & harfbuzz & cb47dca \\
jsoncpp       & 8190e06 & lcms          & f0d9632 & libjpeg        & 3b19db4 & libpcap       & 17ff63e \\
libpng        & cd0ea2a & libxml      & c7260a4 & libxslt        & 180cdb8 & zlib          & d71dc66 \\
openh264*     & 045aeac & openssl*      & b0593c0 & openthread     & 2550699 & vorbis        & 84c0236 \\
\midrule
\multicolumn{2}{c|}{\textbf{Standalone Targets}} & tcpdump       & tcpdump-4.99.4 & mp3gain       & 1.5.2 & jhead          & 3.00 \\ \cline{1-2}
mp42aac       & 1.5.1-628 & xmllint       & libxml2-2.9.14 & bsdtar         & libarchive-3.6.2 & exiv2    & exiv2-0.28.0 \\
ffmpeg        & ffmpeg-6.1 & jq           & jq-1.5 & nm-new         & binutils-2.34 & objdump       & binutils-2.34 \\
pdftotext     & xpdf-4.04 & readelf       & binutils-2.34 & size           & binutils-2.34 & strip-new     & binutils-2.34 \\
\bottomrule
\end{tabular}
\end{table}

\phantomsection
\label{para:exp_setup} 
\textbf{Experiment Setup}. We run  our evaluations on 4 64-bit machines running Ubuntu 22.04 with AMD EPYC™ Genoa 9T24 (128 cores in total) and bound each fuzzing instance to 1 CPU core. Following standard practices in
fuzzing research~\cite{klees2018evaluating}, we perform 10 independent runs for each experiment to ensure statistical significance and each fuzzing campaign runs for 24 hours, which is the standard duration in fuzzing research for obtaining meaningful results. We adhere to standard operating procedures in fuzzing evaluations by binding each fuzzer to a single CPU core. For target programs, we use corpus with small well-formed files as seeds. \added{For our statistical analysis, in addition to reporting standard descriptive metrics such as the mean and standard deviation, we employed the Mann-Whitney U test to confirm the statistical significance of the observed performance differences. This method is extensively adopted in software testing literature for evaluating randomized algorithms and fuzzers~\cite{tests2011arcuri,klees2018evaluating, yu2022htfuzz,zhang2024predecessor, liu2024afgen, zhangunderstanding}, primarily because it is a non-parametric test that operates without relying on assumptions about the underlying data distribution.}

\begin{figure*}[!]
    \centering
    \includegraphics[width=0.85\linewidth]{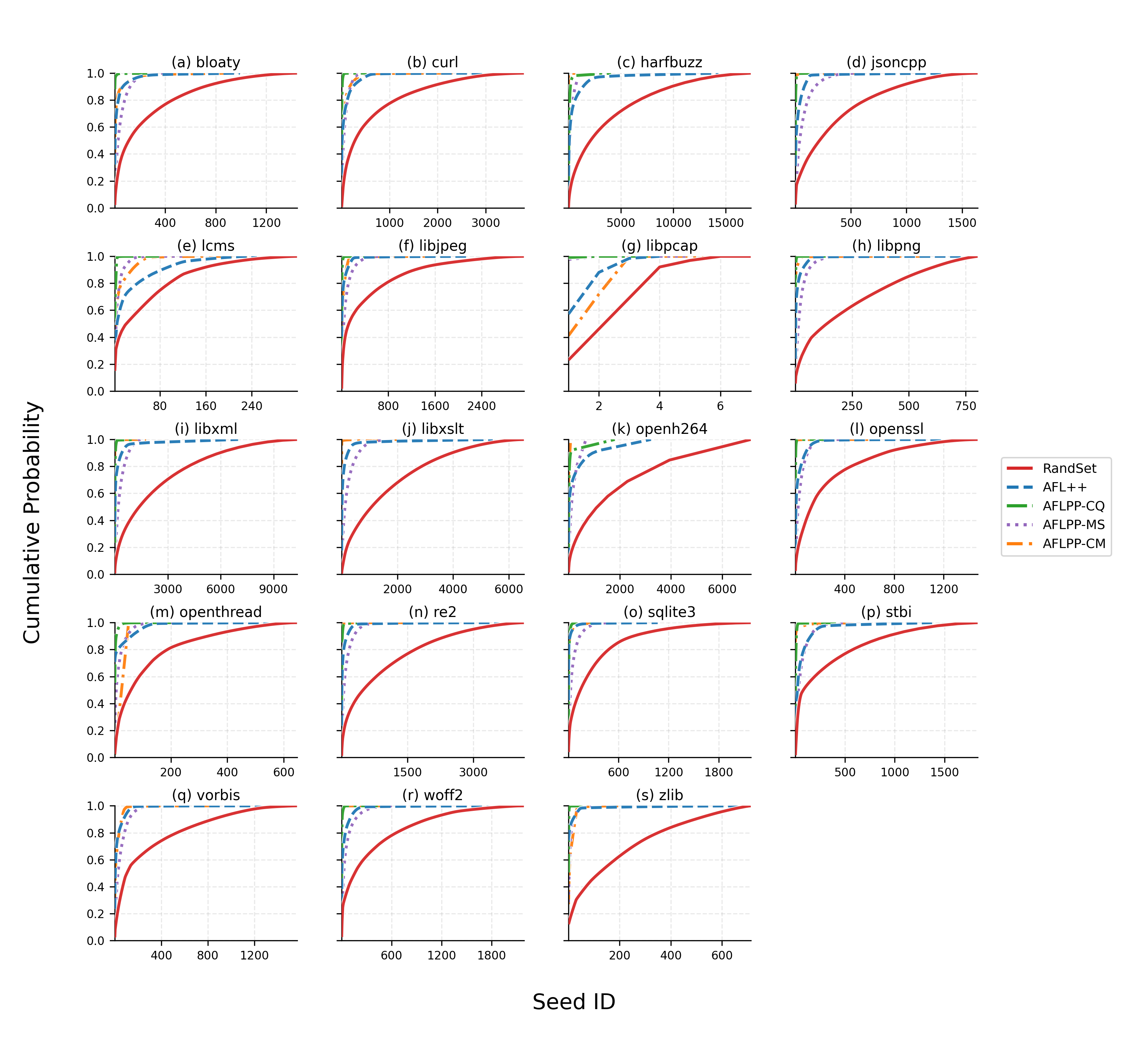}
    \caption{\small \textbf{Cumulative frequency distributions of seeds for \sys against other four baselines on FuzzBench programs over \replaced{10 fuzzing campaigns of 24 hours each}{a 24-hour fuzzing campaign}. The x-axis represents the seed ID sorted in descending order by frequency, the y-axis indicates the cumulative probability of seed selection.}}
    \label{fig:fuzzbench_diversity}
\end{figure*}

\begin{table}[!htbp]
\footnotesize
    \centering
    \footnotesize
    \caption{\small\textbf{Total unique seeds for RandSet and four AFL++-based baselines on 19 FuzzBench programs over \replaced{10 fuzzing campaigns of 24 hours each}{a 24-hour fuzzing campaign}. The highest value for each target is in bold.}}
    {
    \begin{tabular}{lrrrrr}
    \toprule
    \textbf{Targets} & \textbf{\aflpp} & \textbf{\aflppminset} & \textbf{\aflppcq} & \textbf{\aflppcmin} & \textbf{\sys} \\
    \midrule
    bloaty      & 991   & 456  & 383  & 927   & \textbf{1443} \\
    curl        & 3246  & 556  & 524  & 2325  & \textbf{3796} \\
    harfbuzz    & 14290 & 1000 & 4052 & 1751  & \textbf{17377} \\
    jsoncpp     & 1495  & 586  & 192  & 608   & \textbf{1640} \\
    lcms        & 249   & 120  & 78   & 232   & \textbf{319}  \\
    libjpeg     & 2204  & 813  & 503  & 1532  & \textbf{3123} \\
    libpcap     & 4     & 5    & 4    & 6    & \textbf{7}    \\
    libpng      & 778   & 293  & 134  & 604  & \textbf{802}  \\
    libxml      & 6985  & 1436 & 1225 & 1776  & \textbf{10325} \\
    libxslt     & 5425  & 1774 & 1344    & 2059  & \textbf{6540} \\
    openh264    & 3218  & 682  & 1805 & 83  & \textbf{7117} \\
    openssl     & 1229  & 416  & 248  & 611  & \textbf{1474} \\
    openthread  & 640   & 208  & 230  & 50    & \textbf{647}  \\
    re2         & 3594  & 874  & 534  & 1455  & \textbf{4147} \\
    sqlite3     & 1069  & 522  & 346  & 530   & \textbf{2184} \\
    stbi        & 1375  & 573  & 409  & 652  & \textbf{1833} \\
    vorbis      & 1516  & 434  & 302    & 897  & \textbf{1565} \\
    woff2       & 2027  & 726  & 435  & 710  & \textbf{2189} \\
    zlib        & 689   & 130  & 74   & 217   & \textbf{712}  \\
    \bottomrule
    \end{tabular}
    }
    \label{tab:unique_seed_fuzzbench}
\end{table}


\subsection{RQ1: Seed Diversity}
\label{subsec:diversity}
\noindent\textbf{Evaluation Metric.}
To comprehensively evaluate seed diversity of RandSet compared to the other four baselines during whole fuzzing process, we employ two metrics—one qualitative and one quantitative:

1) \textbf{\textit{Cumulative Distribution Frequency (CDF) Plot}}. We use CDF plot to intuitively compare the seed diversity of RandSet and four baselines for each target. In this plot, the x-axis represents seed IDs sorted in descending order according to their selection frequency during fuzzing, while the y-axis shows the cumulative frequency of seed selection. Each point (x, y) on the curve indicates that the top x seeds (ranked by selection frequency) collectively account for a cumulative fraction y of all seed selections. If the curve rises sharply and quickly reaches a high cumulative frequency with only a few seeds (i.e., is steep and concentrated on the left), it means that a small number of seeds dominate the scheduling rounds, reflecting poor diversity. Conversely, a curve that rises more gradually and extends toward the upper right indicates that many different seeds are being selected, demonstrating better diversity.

2) \textbf{\textit{Total Unique Seeds}}. We count the total number of unique seeds that are selected for mutation during the fuzzing process to quantitatively measure seed diversity. This metric directly reflects how broadly the fuzzing engine explores the seed space. A larger value indicates that a wider variety of seeds are being utilized, reflecting better seed diversity. Conversely, a smaller value suggests that only a limited set of seeds are selected, indicating poor diversity.

\added{It is worth noting that the distribution trends and seed characteristics were highly consistent across all 10 trials. Therefore, we randomly selected a single trial to illustrate the trend as it is impractical to present the results of all 10 independent trials simultaneously.}

\textbf{FuzzBench Targets.}
\autoref{fig:fuzzbench_diversity} presents the cumulative selection frequency distributions of RandSet and the four baseline methods on FuzzBench across various targets. As shown, the \sys curves consistently lie at the bottom and exhibit a gradual, smooth ascent from the lower left to the upper right across all targets. In contrast, the baseline curves are concentrated on the left and increase steeply, indicating that seed selection is dominated by a small, fixed subset of high-frequency seeds. the baselines rely on only a few high-frequency seeds, whereas \sys requires a significantly larger number of seeds. 

The total unique seeds results on FuzzBench are presented in \autoref{tab:unique_seed_fuzzbench}. It is evident that \sys outperforms all other baselines across all FuzzBench targets. For example, on the libxml target, \sys explores 10,325 unique seeds, whereas \aflpp explores only 6,985, \aflppminset 1,436, \aflppcq 1,225, and \aflppcmin 1,776. This clearly demonstrates that RandSet enables the fuzzing engine to more broadly and thoroughly utilize the seed corpus. Taken together, the results of these two metrics consistently show that \sys achieves superior seed diversity compared to all baselines, allowing for more comprehensive exploration of the seed space and effectively preventing computational resources from being concentrated on a small, fixed set of seeds.

\textbf{Standalone Targets.}
The cumulative selection frequency distributions and total unique seeds for standalone targets are presented in \autoref{fig:standalone_diversity} and \autoref{tab:unique_seed_standalone}, respectively. We observed similar trends to those on FuzzBench.

In terms of total unique seeds, \sys outperforms the baselines on 9 out of 15 targets. A representative target is \texttt{jq}, in which \sys explores 1,958 unique seeds, whereas other four baselines only discover 833, 497, 352, and 58 unique seeds, respectively. Regarding targets where \sys loses to the best baseline (e.g., tcpdump), this is largely because standalone programs tend to be larger and slower, thus needing more time to hit the coverage plateau. Notably, the advantage of our \sys becomes more apparent especially when fuzzing gets stuck, as the seed corpus stabilizes and changes very little. In such scenarios, other deterministic corpus reduction methods almost always produce a small, fixed subset of seeds, causing the fuzzing engine to continually mutate the same set of seeds, which is highly inefficient. In contrast, \sys effectively mitigates this issue by maintaining better seed diversity.

\begin{longfbox}
\textbf{Result 1:} 
The cumulative frequency distribution plots demonstrate that \sys consistently maintains a smoother and slower-growing curve, indicating superior seed diversity, in contrast to the sharp rises observed for the baseline methods on both standalone and FuzzBench targets. Meanwhile, \sys explores more unique seeds than all the baselines on 28 out of 34 targets.
\end{longfbox}

\subsection{RQ2: Corpus Reduction}
\label{reduction}
\noindent\textbf{Evaluation Metric.}
To answer RQ2, we use the subset ratio as a unified evaluation metric to assess the corpus reduction effectiveness of \sys compared to all four baselines. Specifically, for each method, we record both the final corpus size and the size of the last constructed subset at the end of the 24-hour fuzzing campaign. Specifically, for \aflppcq, We consider the seed set marked as \textit{favored} to be the selected subset. As for \aflppcmin, we take the seeds retained after \texttt{afl-cmin} minimization as the subset. Besides, it is worth noting that there is no concept of subset in vanilla \aflpp, because seeds that are not marked as \textit{favored} may also be selected, so we do not consider this baseline for this metric.

\begin{table}[!htbp]
\scriptsize
    \centering
    \caption{\small\textbf{Corpus subset ratio (\%) of \sys compared to three baselines on FuzzBench targets after 24-hour fuzzing campaigns. We mark the smallest subset ratio for each target in bold.}}
    \resizebox{\columnwidth}{!}{
    \begin{tabular}{lrrrr|lrrrr}
    \toprule
    \textbf{Targets} & \textbf{\aflppminset} & \textbf{\aflppcq} & \textbf{\aflppcmin} & \textbf{\sys} & \textbf{Targets} & \textbf{\aflppminset} & \textbf{\aflppcq} & \textbf{\aflppcmin} & \textbf{\sys} \\
    \midrule
    bloaty              & 12.85  & 21.22  & 46.00 & \textbf{4.17}  & openssl             & 7.72   & 6.52   & 49.62 & \textbf{0.79}  \\
    curl                & 18.02  & 19.17  & 48.18 & \textbf{4.85}  & openthread          & 11.82  & 16.69  & 38.71 & \textbf{2.85}  \\
    harfbuzz            & 14.07  & 12.95  & 45.06 & \textbf{2.13}  & re2                 & 15.17  & 19.32  & 35.73 & \textbf{3.66}  \\
    jsoncpp             & 5.73   & 9.36   & 34.12 & \textbf{2.47}  & sqlite3             & 6.10   & 8.16   & 42.42 & \textbf{2.25}  \\
    lcms                & 28.07  & 31.77  & 56.86 & \textbf{4.91}  & stbi                & 4.72   & 6.36   & 46.80 & \textbf{1.14}  \\
    libjpeg             & 5.51   & 8.30   & 35.93 & \textbf{2.00}  & woff2               & 9.15   & 10.98  & 37.89 & \textbf{3.51}  \\
    libpcap             & 100.00 & 100.00 & 100.00 & \textbf{70.37} & zlib                & 9.49   & 10.52  & 18.88 & \textbf{1.01}  \\
    libpng              & 5.87   & 11.55  & 41.15 & \textbf{2.29}  & libxslt             & 8.53   & 9.86   & 33.10 & \textbf{1.66}  \\
    libxml              & 4.95   & 6.96   & 31.85 & \textbf{1.63}  & vorbis              & 5.36   & 6.09   & 39.51 & \textbf{0.90}  \\ \cline{6-10}
    openh264            & 3.96   & 4.97   & 35.03 & \textbf{1.22}  &          \textbf{Mean Ratio}            &   \textbf{14.58}     &   \textbf{16.88}     &   \textbf{42.99}   &   \textbf{5.99}             \\
    \bottomrule
    \end{tabular}
    }
    \label{tab:fuzzbench_reduction}
\end{table}

\begin{table}[!htbp]
\small
    \centering
    \caption{\small\textbf{Corpus subset ratio (\%) of \sys compared to three baselines on standalone targets after 24-hour fuzzing campaigns. We mark the smallest subset ratio for each target in bold.}}
    \resizebox{\columnwidth}{!}{
    \begin{tabular}{lrrrr|lrrrr}
    \toprule
    \textbf{Targets} & \textbf{\aflppminset} & \textbf{\aflppcq} & \textbf{\aflppcmin} & \textbf{\sys} & \textbf{Targets} & \textbf{\aflppminset} & \textbf{\aflppcq} & \textbf{\aflppcmin} & \textbf{\sys} \\
    \midrule
    bsdtar      & 8.91   & 12.98  & 44.14 & \textbf{3.06} & objdump     & 11.42  & 15.95  & 52.55 & \textbf{3.78} \\
    exiv2       & 12.60  & 12.96  & 52.25 & \textbf{8.37} & pdftotext   & 6.72   & 11.00  & 45.42 & \textbf{1.12} \\
    ffmpeg      & 10.93  & 10.04  & 53.26 & \textbf{3.33} & readelf     & 19.14  & 17.85  & 72.11 & \textbf{1.21} \\
    jhead       & 6.91   & 10.14  & 38.29 & \textbf{1.50} & size        & 17.59  & 25.35  & 51.38 & \textbf{8.08} \\
    jq          & 4.17   & 7.07   & 37.73 & \textbf{1.76} & strip-new   & 19.58  & 23.04  & 37.12 & \textbf{7.69} \\
    mp3gain     & 3.01   & 3.92   & 29.23 & \textbf{1.04} & tcpdump     & 26.34  & 27.77  & 61.18 & \textbf{9.35} \\
    mp42aac     & 13.45  & 20.15  & 44.78 & \textbf{2.71} & xmllint     & 5.92   & 8.19   & 30.04 & \textbf{2.71} \\ \cline{6-10}
    nm-new      & 17.42  & 23.92  & 60.11 & \textbf{4.78} & \textbf{Mean Ratio} & \textbf{12.27} & \textbf{15.35} & \textbf{47.31} & \textbf{4.03} \\
    \bottomrule
    \end{tabular}
    }
    \label{tab:standalone_reduction}
\end{table}

 \autoref{tab:fuzzbench_reduction} and \autoref{tab:standalone_reduction} present the final subset ratios of each method on FuzzBench and standalone targets, respectively. \sys consistently achieves the lowest subset ratio across all targets, with remarkable averages of 5.99\% and 4.03\%, while significantly better than all other baselines. A notable example is the \texttt{readelf} target, where \sys achieves a subset ratio of only 1.21\% of its original size while preserving full feature coverage, indicating substantial redundancy in the original corpus. In contrast, the other three baselines exhibit much higher subset ratios: 19.14\% for \aflppminset, 17.85\% for \aflppcq and 72.11\% for \aflppcmin. As for target \texttt{libpcap}, \sys yields a relatively high subset ratio of 70.37\%. The main reason is that this target has a very small total number of covered edges, only around 50, which means the overall size of seed corpus is very small, leading to a higher subset ratio. In fact, the subset ratios of the other three baselines on this target are all 100\%.

The results demonstrate that our \sys is more effective in corpus reduction than existing baseline approaches. This superiority primarily stems from \sys’s adoption of frontier nodes, rather than edges or basic blocks, as the
feature in the set cover formulation. Frontier node provides a more expressive and fine-grained representation of program coverage, resulting in a reduced feature space that contributes to \sys’s smaller subset ratio.


\begin{longfbox}
\textbf{Result 2:} 
\sys achieves impressive average subset ratios across all FuzzBench and standalone targets, with averages of 5.99\% and 4.03\% , respectively.
\end{longfbox}

\subsection{RQ3: Code Coverage}
\label{subsec:codecov}
To answer RQ3, we evaluate the code coverage achieved by \sys against all four baselines. We test across both \numfuzzbench FuzzBench and \numstandalone standalone targets. All experiments consist of 10 independent 24-hour campaigns. Coverage is measured using the built-in coverage collection modules in each fuzzing framework. \added{The Mann-Whitney U test results are reported in \autoref{sec:appendix_mwu}.}

\textbf{Standalone Targets.}
The coverage comparison between \sys and these four baselines is shown in \autoref{tab:standalone}. Among the \numstandalone evaluated standalone programs, \sys outperforms all baselines on 10 targets, achieving on average 16.58\% higher code coverage compared to vanilla \aflpp, 9.49\% higher than \aflppminset, 17.52\% higher than \aflppcq, and 50.54\% higher than \aflppcmin across the standalone target set. A representative example is \texttt{exiv2}, where \sys discovers 124.7\% more edges compared to AFL++, 35.2\% more than \aflppminset, 38.4\% more than \aflppcq, and 112.0\% more than \aflppcmin. 

For standalone programs that are typically larger and slower to execute compared to FuzzBench targets, seed scheduling becomes particularly important. \sys excels in this scenario by maintaining a small but high-quality corpus subset to assist seed scheduling. This approach enables \sys to efficiently select seeds that are most worth exploring within the limited time budget, while ensuring coverage across diverse program features.

\begin{table}[!htbp]
\small
\setlength{\tabcolsep}{2pt} 
    \centering
    \caption{\small\textbf{Mean edge coverage of RandSet against AFL++-based baselines on FuzzBench programs for 24 hours over 10 runs. \replaced{We mark the highest number in bold.}{Results are presented as Mean $\pm$ Std, with the highest numbers marked in bold.}}}
    {
    \begin{tabular}{l *{5}{r @{\,{\scriptsize$\pm$}\,} >{\scriptsize}l}}
    \toprule
    \textbf{Targets} & 
    \multicolumn{2}{c}{\textbf{AFL++}} & 
    \multicolumn{2}{c}{\textbf{\aflppminset}} & 
    \multicolumn{2}{c}{\textbf{\aflppcq}} & 
    \multicolumn{2}{c}{\textbf{\aflppcmin}} & 
    \multicolumn{2}{c}{\textbf{\sys}} \\
    \midrule
    bloaty       & 3101.3 & 92   & 3408 & 37      & 2913 & 119      & 3103.5 & 152        & \textbf{3429.4} & 555 \\
    curl         & 14238.6 & 439 & 12603.4 & 519  & 11846 & 598     & 13403.2 & 507    & \textbf{14655.6} & 684 \\
    harfbuzz     & 23682.7 & 435 & 17749.4 & 926  & 21939.4 & 566   & 17094.0 & 1550   & \textbf{24045} & 1019 \\
    jsoncpp      & 1337.1 & 2    & 1339.6 & 9     & 1332.4 & 12     & 1333.1 & 3       & \textbf{1341.4} & 0 \\
    lcms         & \textbf{1518.3} & 450  & 1466 & 310     & 986 & 85        & 1165.6 & 352       & 1404.4 & 482 \\
    libjpeg      & 3232.5 & 21   & 2999.2 & 6     & 3017.2 & 250    & \textbf{3251.5} & 214     & 3241.67 & 20 \\
    libpcap      & 49.2 & 5      & 43 & 0         & 43 & 4          & 51 & 0           & \textbf{51} & 4 \\
    libpng       & \textbf{1382.4} & 99   & 1360.4 & 0     & 1360.2 & 23     & 1360.4 & 99      & 1361 & 0 \\
    libxml       & 9754.4 & 996  & 8530.2 & 520   & 10323.4 & 1945  & 8735 & 346       & \textbf{13124.4} & 672 \\
    libxslt      & \textbf{4699.8} & 20   & 4551.33 & 92   & 4648 & 82       & 4655.5 & 130     & 4631.2 & 58 \\
    openh264     & 13664.2 & 218 & 13645.4 & 209  & \textbf{13944.2} & 38    & 11465.3 & 5302   & 13824.2 & 93 \\
    openssl      & 4641 & 54     & 4535 & 43      & 4429 & 69       & 4501.3 & 122     & \textbf{4727.2} & 46 \\
    openthread   & 4925.4 & 456  & 4981.6 & 169   & 4781.6 & 132    & 4180.2 & 439     & \textbf{5034} & 93 \\
    re2          & 6252.9 & 60   & 6260.2 & 53    & 6251.2 & 40     & 6255.4 & 52      & \textbf{6270.8} & 14 \\
    sqlite3      & 11941.2 & 216 & 11866.4 & 245  & 11717.8 & 83    & 11557.4 & 178    & \textbf{12116.6} & 100 \\
    stbi         & 2331.4 & 216  & 2788 & 271     & 2308 & 351      & 2357.5 & 223     & \textbf{2798.3} & 132 \\
    vorbis       & \textbf{2051.9} & 22   & 2041.7 & 4     & 2040.0 & 6      & 2037.5 & 33      & 2042.7 & 9 \\
    woff2        & 2425.9 & 27   & 2380.8 & 53    & 2406.2 & 85     & 2388.2 & 54      & \textbf{2458.5} & 40 \\
    zlib         & 879.1 & 7     & 888.4 & 2      & 879.4 & 7       & \textbf{897.0} & 8      & 882.6 & 4 \\
    \midrule
    \textbf{Mean Gain} & 
    \multicolumn{2}{c}{\textbf{3.57\%}} & 
    \multicolumn{2}{c}{\textbf{7.28\%}} & 
    \multicolumn{2}{c}{\textbf{9.55\%}} & 
    \multicolumn{2}{c}{\textbf{10.46\%}} & 
    \multicolumn{2}{c}{—}  \\
    \bottomrule
    \end{tabular}
    }
    \label{eval:fuzzbench_aflpp}
\end{table}

\textbf{FuzzBench Targets.}
Across FuzzBench targets (see results in \autoref{eval:fuzzbench_aflpp}), \sys outperforms all four baselines on 12 programs, achieving an average coverage increase of 3.57\% over vanilla \aflpp, 7.28\% over \aflppminset, 9.55\% over \aflppcq, and 10.46\% over \aflppcmin. This improvement over vanilla \aflpp is particularly significant given that many FuzzBench targets are relatively small programs where achieving substantial coverage differences is challenging, and recent fuzzing works\cite{she2024fox, xie2025ztaint, zheng2025mendelfuzz} typically report only modest coverage improvements on these benchmarks. A representative example is \texttt{libxml}, where our method covers 34.6\%, 53.9\%, 27.1\% and 50.3\% more edges than the other four baselines, respectively. 
\added{This superior performance demonstrates that for targets with large program spaces, \sys's ability to identify redundant seeds and enhance diversity significantly promotes the efficient exploration of complex program logic. This is evidenced by \autoref{tab:unique_seed_fuzzbench} and \autoref{tab:fuzzbench_reduction}, which show that while achieving an effective corpus reduction rate of 1.63\%, \sys selected 10,325 unique seeds for mutation. In contrast, the other baselines selected only 6,985, 1,436, 1,225, and 1,776  seeds, respectively.}
\replaced{For a few targets, such as \texttt{lcms} and \texttt{libpng}, \sys achieves slightly lower coverage compared to the highest coverage recorded by any of the baselines for those targets. This can be attributed to these targets’ heavy reliance on constraint solving capabilities—when the baseline fuzzer has limited constraint solving ability, occasional successful constraint solving in a single instance may lead to apparent performance gaps; however, these differences are not statistically significant across multiple runs.}{For a few targets, such as \texttt{lcms} and \texttt{libpng}, \sys achieves slightly lower coverage than the best baseline result. This discrepancy can be attributed to outlier runs on constraint-heavy targets. Since significant coverage gains on these targets often depend on solving a few complex constraints, the baseline fuzzer may occasionally achieve a breakthrough in a single run.  For instance, examining the coverage data of \texttt{lcms} across 10 runs reveals that vanilla AFL++ surpasses our method in only 2 out of 10 runs, both characterized by sudden spikes. By further mapping coverage to the source code, we confirmed that these spikes arise because vanilla AFL++ happened to solve a few additional complex constraints during these specific runs.}

Particularly noteworthy is \sys's ability to reduce the input corpus into a compact yet effective subset, enabling more efficient seed scheduling. This effect is elaborated on in \autoref{reduction}. Through our corpus reduction strategy, \sys maintains a minimal set of high-quality seeds that achieve same feature coverage to the full corpus, which directly contributes to the consistent performance improvements across all baselines.

\begin{longfbox}
\textbf{Result 3:} 
\sys outperforms all four baselines. On standalone targets, it achieves average coverage improvements of 16.58\% over vanilla \aflpp, 9.49\% over \aflppminset, 17.52\% over \aflppcq, and 50.54\% over \aflppcmin. On FuzzBench targets, the respective improvements are 3.57\%, 7.28\%, 9.55\%, and 10.46\%.
\end{longfbox}

\subsection{RQ4: Bug Discovery}

\replaced{To answer RQ4, we assess the bug discovery capabilities of \sys by comparing it with three baselines, as the \aflppcmin baseline is not compatible with the Magma setup. Following the guidelines proposed by Klees et al.~\cite{klees2018evaluating} for evaluating fuzzers using ground-truth bugs, we conducted experiments on the Magma benchmark~\cite{hazimeh2020magma}. Specifically, for each method, we record both the cumulative number of unique bugs triggered and reached during the 24-hour fuzzing campaign over 5 runs to evaluate the effectiveness of bug detection across different approaches.}{To answer RQ4, we assess the bug discovery capabilities of \sys on the Magma benchmark~\cite{hazimeh2020magma} and real-world programs (including both FuzzBench and standalone targets). We compare \sys against all four baselines, with the exception of Magma where \aflppcmin is excluded due to incompatibility.}


\textbf {Magma Dataset.}
\added{Following the guidelines proposed by Klees et al.~\cite{klees2018evaluating} for evaluating fuzzers using ground-truth bugs,} we evaluated our approach using Magma, a widely-adopted benchmark in the fuzzing community~\cite{lee2023learning,herrera2022dataflow, fang2024ddgf, zhang2025low}. Magma comprises 21 programs from nine open-source libraries with known, deliberately injected vulnerabilities. Our evaluation covered 17 programs from eight libraries: \texttt{libpng}, \texttt{libtiff}, \texttt{libxml2}, \texttt{lua}, \texttt{openssl}, \texttt{poppler}, \texttt{sqlite3}, and \texttt{libsndfile}. We excluded \texttt{php} and its four associated fuzz drivers due to dependency conflicts during the final linking stage. Specifically, for each baseline, we record the cumulative number of unique bugs triggered and reached on Magma during the 24-hour fuzzing campaign over 5 independent runs to evaluate the effectiveness of bug detection.

\begin{table}[!htbp]
\scriptsize
    \centering
    \caption{\small\textbf{Cumulative number of unique bugs triggered and reached of \sys against all three baselines in Magma programs for 24 hours over 5 runs. (triggered $|$ reached)}}
    {
    \begin{tabular}{lcccc}
    \toprule
    \textbf{Targets} & \textbf{AFL++} & \textbf{\aflppminset} & \textbf{\aflppcq} & \textbf{\sys} \\
    \midrule
    asn1            & 2$|$4   & 2$|$4   & 2$|$4   & 2$|$4   \\
    asn1parse       & 0$|$1   & 0$|$1   & 0$|$1   & 0$|$1   \\
    bignum          & 0$|$1   & 0$|$1   & 0$|$1   & 0$|$1   \\
    client          & 1$|$7   & 1$|$7   & 1$|$7   & 1$|$7   \\
    libpng          & 2$|$6   & 2$|$6   & 2$|$6   & 2$|$6   \\
    libxml2\_xml    & 4$|$8   & 4$|$8   & 4$|$8   & 5$|$9   \\
    lua             & 1$|$4   & 1$|$2   & 2$|$3   & 2$|$4   \\
    pdf\_fuzzer     & 3$|$16  & 3$|$16  & 2$|$15  & 3$|$17  \\
    pdfimages       & 3$|$12  & 2$|$12  & 3$|$13  & 4$|$13  \\
    pdftoppm        & 3$|$16  & 2$|$15  & 1$|$15  & 3$|$16  \\
    server          & 1$|$6   & 1$|$6   & 1$|$6   & 1$|$6   \\
    sqlite3\_fuzz   & 3$|$12  & 3$|$10  & 3$|$10  & 5$|$13  \\
    sndfile\_fuzzer & 7$|$8   & 7$|$8   & 7$|$8   & 7$|$8   \\
    tiff\_read      & 3$|$7   & 3$|$5   & 3$|$5   & 4$|$7   \\
    tiffcp          & 4$|$6   & 3$|$5   & 4$|$7   & 4$|$6   \\
    xmllint         & 3$|$7   & 3$|$7   & 3$|$7   & 4$|$8   \\
    x509            & 1$|$5   & 1$|$5   & 1$|$5   & 1$|$5   \\
    \midrule
    \textbf{total}  & \textbf{41$|$126} & \textbf{38$|$118} & \textbf{39$|$121} & \textbf{48$|$131} \\
    \bottomrule
    \end{tabular}
    }
    \scriptsize
    \label{tab:magma}
\end{table}


The results, detailed in \autoref{tab:magma}, demonstrate that \sys consistently outperforms all three baselines in bug discovery. Specifically, \sys triggers 48 unique bugs compared to 41, 38, and 39 bugs triggered by AFL++, \aflppminset, and \aflppcq respectively, representing a 17.1\% improvement over vanilla AFL++. In terms of bugs reached, \sys achieves 131 compared to 126, 118, and 121 for the three baselines. Notably, \sys shows superior performance on several challenging targets such as \texttt{libxml2\_xml}, \texttt{sqlite3\_fuzz}, and \texttt{xmllint}, where it triggers more bugs than any baseline. This improvement in bug-finding capability aligns with Bohme et al.'s~\cite{bohme2022reliability} observation that there exists a strong correlation between code coverage and bug discovery.

\added{
\textbf {Real-world Bugs}
To ensure the generality of \sys's bug-finding capabilities across broader domains, we extended our evaluation to real-world programs, including both FuzzBench and standalone targets. In this assessment, we recorded the cumulative number of unique crashes triggered by each baseline throughout the experiments described in Section \ref{para:exp_setup}.

As presented in \autoref{tab:realwrold_bugs}, \replaced{Randset}{\sys} achieved the best performance by triggering 21 unique bugs, compared to 20, 16, 18, and 5 bugs found by \aflpp, \aflppminset, \aflppcq, and \aflppcmin, respectively. It is worth noting that \sys successfully triggered a bug in the woff2 target that was missed by all other baselines, indicating that \replaced{Randset}{\sys} exhibits superior performance in real-world bug discovery as well.
}

\begin{longfbox}
\textbf{Result 4:} 
\sys triggers up to \maxmagma more ground-truth bugs on Magma \added{and 1 more bug on real-world programs} compared to the state-of-the-art fuzzer.
\end{longfbox}

\subsection{RQ5: \replaced{Runtime Efficiency}{Performance Overhead}}
\noindent\textbf{Evaluation Metric.}
\replaced{To evaluate the runtime overhead incurred by \sys for corpus reduction, we calculate the proportion of the total fuzzing time spent on subset generation. This metric is compared across \sys and three baselines: \aflppminset, \aflppcq, and \aflppcmin. Notably, as analyzed in \autoref{tab:cull_vs_random}, the \texttt{cull\_queue} in vanilla AFL++ is barely used during fuzzing, so we do not measure this metric for the vanilla \aflpp baseline.}{To evaluate the performance overhead of \sys, we measure both the one-time offline cost of CFG construction and the runtime overhead of subset generation. Regarding the runtime component, we calculate the percentage of total fuzzing time consumed by subset generation, comparing \sys against three baselines: \aflppminset, \aflppcq, and \aflppcmin. Notably, as
analyzed in Table 2, the \texttt{cull\_queue} in vanilla AFL++ is barely used during fuzzing, so we do not measure this metric for the vanilla AFL++ baseline. Furthermore, we decompose the runtime cost to isolate the specific cost of frontier node calculation.}

\added{\textbf {Runtime Overhead.}}
\autoref{tab:fuzzbench_overhead} and \autoref{tab:standalone_overhead} present the runtime overhead of \sys and the three baseline corpus reduction techniques on FuzzBench and standalone targets. As shown, \sys spends only 1.17\% of the total fuzzing time on subset generation for FuzzBench targets, and only 3.93\% for standalone targets, both of which are considered acceptable in practice. In contrast, both \aflppminset and \aflppcmin introduce substantial overhead. On standalone targets, their average overheads reach 22.23\% and 14.93\%, respectively. On FuzzBench targets, \aflppcmin incurs 9.93\% overhead,while \aflppminset reaches a remarkable 56.08\%, meaning more than half of the fuzzing time is spent on this corpus reduction algorithm, which is highly inefficient. This result aligns with our analysis in \autoref{limit_of_existing_tech}. The high overhead of \aflppminset arises from its greedy set cover algorithm, which requires repeatedly scanning the entire seed corpus each time a subset is selected. For \aflppcmin, the primary source of overhead stems from expensive disk I/O operations. As for \aflppcq, it achieves the lowest overhead among all baselines, which is expected. The greedy algorithm in \texttt{cull\_queue} is not equivalent to a traditional greedy set cover algorithm; instead of selecting seeds that cover the most edges, it chooses seeds that execute fastest and are smallest in size, thereby avoiding repeated traversals of the seed queue. While our randomized set cover algorithm introduces some performance penalty compared to the deterministic greedy approach in \texttt{cull\_queue}, this trade-off is justified: \sys achieves significantly better seed diversity and more thorough corpus reduction.

\added{\autoref{tab:aflpp_frontier_overhead} breaks down the runtime overhead introduced by \sys's frontier node computation for AFL++ integrations on both FuzzBench  and standalone targets. The results indicate that the runtime overhead is minimal, accounting for less than 0.01\% of total fuzzing time across almost all targets. This is expected because the time complexity for calculating the frontier node for each seed is \(O(n)\), where n is the bitmap size. Consistent with findings in \cite{ahmed2021bigmap}, bitmap operations are nearly free when n is small. Although large-bitmap target such as \texttt{ffmpeg} show a higher relative overhead, 0.026\%, its lower throughput reduces the number of times the bitmap is scanned, so the overall overhead remains small. Therefore, the runtime overhead introduced by \sys's frontier node computation is negligible in practice.}

\added{\textbf {Offline Overhead.}}
\added{As described in Section \autoref{cfg_construction_description}, we utilize a modified LLVM pass to extract basic block-level intra-procedural CFG information at compile-time and store it within a custom section of the target binary. A Python script is then employed to read this section offline and generate an external CFG dictionary. \autoref{tab:aflpp_cfg_time} presents the time overhead of this procedure across all FuzzBench and standalone targets. In general, the overhead is lightweight, with the majority of targets requiring less than one minute. Even for ffmpeg, the most time-consuming target, the process took only about 4 minutes. Given that this is a one-time offline effort, the overhead is considered acceptable.}

\begin{table}[!ht]
\scriptsize
    \centering    \caption{\small\textbf{\added{Runtime} overhead (\%) of \sys compared to three baselines on FuzzBench targets, \added{measured as the percentage of total fuzzing time}. Mean Ratio is the average overhead across all targets.}}
    \begin{tabular}{lrrrr}
    \toprule
    \textbf{Targets} & \textbf{\aflppminset} & \textbf{\aflppcq} & \textbf{\aflppcmin} & \textbf{\sys} \\
    \midrule
    bloaty             & 37.16  & 0.10   & 4.52  & 0.13 \\
    curl                    & 96.25  & 6.46   & 6.82  & 3.96 \\
    harfbuzz                & 98.40  & 5.04   & 11.99  & 7.63 \\
    jsoncpp                 & 4.16   & 0.04   & 15.89 & 0.08 \\
    lcms                    & 1.48   & 0.02   & 11.54 & 0.04 \\
    libjpeg                 & 83.97  & 0.26   & 24.66  & 0.87 \\
    libpcap                 & 0.01   & 0.08   & 0.06  & 0.04 \\
    libpng             & 28.64  & 0.12   & 11.84 & 0.21 \\
    libxml                  & 97.16  & 1.85   & 10.17 & 3.85 \\
    openh264            & 92.14  & 0.42   & 0.23  & 1.17 \\
    openssl                 & 59.29  & 0.21   & 18.26  & 0.12 \\
    openthread      & 76.85  & 1.26   & 28.82 & 0.88 \\
    re2                & 53.36  & 0.16   & 7.65  & 0.23 \\
    sqlite3 & 86.29  & 0.22   & 11.77  & 1.09 \\
    stbi                    & 82.35  & 0.68   & 8.30   & 0.67 \\
    woff2                   & 10.46  & 0.03   & 2.01 & 0.04 \\
    zlib                    & 84.10  & 0.30   & 0.74  & 0.09 \\
    libxslt          & 71.21  & 0.75   & 9.70 & 1.13 \\
    vorbis                  & 2.20   & 0.01   & 3.68  & 0.00 \\
    \midrule
    \textbf{Mean Ratio}     & \textbf{56.08}  & \textbf{0.95}   & \textbf{9.93} & \textbf{1.17} \\
    \bottomrule
    \end{tabular}
    \label{tab:fuzzbench_overhead}
\end{table}

\begin{longfbox}
\textbf{Result 5:} 
The runtime overhead introduced by \sys during corpus reduction is entirely acceptable, with only 1.17\% on FuzzBench targets and 3.93\% on standalone targets. \added{Furthermore, the offline overhead is negligible, requiring less than 1 minute for most targets.}

\end{longfbox}

\subsection{RQ6: Generalizability}
\label{app:general}
To evaluate the generalizability of \sys beyond \aflpp, we integrated \sys into two other state-of-the-art fuzzing engines, \libafl and \centipede, and conducted experiments with them. Each evaluation was performed on the FuzzBench dataset under the recommended configurations. We did not assess these fuzzers on standalone programs, as both \libafl and \centipede are primarily designed for fuzzing libraries and are best suited to the FuzzBench environment. The full list of evaluated targets is shown in \autoref{tab:programs}. We used three key metrics to assess the performance: coverage, subset ratio, and \replaced{runtime overhead}{performance overhead}. \added{The Mann--Whitney U test results are reported in \autoref{sec:appendix_mwu}.}

Our results show that \sys consistently enhances performance across different fuzzing engines. As illustrated in \autoref{eval:fuzzbench_nonaflpp}, \sys-integrated \libafl achieves an average coverage improvement of \maxlibaflfuzzbenchcov over its baseline, while \centipede gains \maxcentipedefuzzbenchcov. \added{Despite these overall improvements, we further analyze cases where \sys underperforms the baseline. For the high-throughput \libafl engine, we hypothesize that \sys's aggressive corpus reduction strategy might hurt targets with naturally small corpora (e.g., \texttt{libpng}, \texttt{vorbis}) by discarding some intermediate seeds that are worth exploring, especially under \libafl's high fuzzing speed. Conversely, for \centipede on \texttt{openh264}, the huge program size and complex parsing logic lead to significantly low fuzzing speed. In this scenario, \sys's diverse seed selection make the limited mutation budget too dispersed and fails to penetrate complex branch constraints.
}

The corpus reduction results, presented in \autoref{tab:fuzzbench_reduction_non_aflpp_fuzzer}, demonstrate that \libafl and \centipede achieve subset ratios of \libaflfuzzbenchdistill\% and \centipedefuzzbenchdistill\%, respectively, with no loss in fuzzing effectiveness.
We also measured the \replaced{runtime overhead}{performance overhead}  introduced by \sys. As shown in \autoref{tab:overheadfuzzbench_non_aflpp_fuzzers}, the additional overhead remains minimal, averaging \libaflfuzzbenchover\% for \libafl and \centipedefuzzbenchover\% for \centipede. \added{\autoref{tab:centipede_libafl_cfg_construction_time} presents the time required to construct the CFG using block-level intra-procedural information from LLVM instrumentation. It is worth noting that Centipede provides native support for this functionality via the \texttt{use\_coverage\_frontier=true} startup flag. As observed, this overhead is negligible (generally less than 1 s) for both LibAFL and Centipede on all targets. Furthermore, \autoref{tab:centipede_libafl_frontier_node_overhead} details the percentage of total fuzzing time spent calculating frontier nodes. The average runtime overhead is 4.211\% for LibAFL and 0.172\% for Centipede.} These results indicate that \sys imposes negligible performance cost while improving corpus efficiency and enhancing coverage.

\begin{longfbox}
\textbf{Result 6:} \sys generalizes well to multiple fuzzing engines, demonstrating consistent gains in coverage and corpus reduction with minimal runtime overhead across \libafl, and \centipede.
\end{longfbox}

\subsection{RQ7: Ablation Study}
\label{app:abalation}
To evaluate the effectiveness of individual components in \sys, We conducted an ablation study in which each of the two core components was disabled individually. Based on our design, we developed two variants in addition to the complete \sys: (i) R-G, which implements a greedy approach for set construction instead of our randomized strategy and (ii) R-E, which we replaced the feature used for set cover from frontier nodes to edges. We measured three metrics including seed diversity, subset ratio and code coverage achieved by these variants on the FuzzBench benchmark through 10 independent one-hour trials.

For each target, we plotted their cumulative frequency distribution curves to illustrate the diversity of \sys compared to other variants. As shown in \autoref{fig:ablatation_diversity}, the R-G variant (green curve) consistently lies to the left of the others and exhibits a markedly steeper ascent. In contrast, the other two variants employing the randomized algorithm show almost identical trends, with smoother and more gradual curves. \autoref{tab:ablatoin_total_unique_seeds} presents the total number of unique seeds for \sys and the other two variants on FuzzBench. It is evident that the R-G variant, which does not incorporate the randomized design, achieves a significantly lower total number of unique seeds compared to the other two variants. These results collectively demonstrate that our randomized algorithm makes a significant contribution to seed diversity, leading to a more even and widespread distribution of seeds.

\autoref{tab:fuzzbenchablation_subset_ratio} shows the final subset ratios of different variants on FuzzBench. We observe that the mean ratios of complete \sys and R-G are similar, all maintaining a relatively low average. In contrast, R-E exhibits a much higher subset ratio—about three times that of the other variants. This result demonstrates the superiority and effectiveness of using the more expressive frontier node as the feature in our set cover formulation. A smaller feature size leads to a lower subset ratio, which in turn indicates more effective corpus reduction. \autoref{tab:fuzzbenchablation_coverage} presents the final average coverage achieved by each variant on FuzzBench. As shown, complete \sys achieves the highest mean coverage among all variants, outperforming R-G by 3.96\% and R-E by 1.94\%. This suggests that each component plays an important role in improving coverage, and that their combination yields the best results.

\begin{longfbox}
\textbf{Result 7:} 
Both components of \sys demonstrate their effectiveness, with the complete version showing better diversity, lower subset ratio and higher coverage on average.
\end{longfbox}

\subsection{\added{RQ8: Alternative Set Cover Algorithms}}
\label{sub:LP_based_rounding}
\added{
In this section, we investigate whether other existing algorithms for the Set Cover problem can serve as effective alternatives to \sys's subset selection strategy for corpus reduction during fuzzing seed scheduling. Specifically, we consider two classic approximation algorithms: randomized rounding and standard rounding \cite{vazirani2001approximation}. Both algorithms start by relaxing the problem into a Linear Programming (LP) model to obtain fractional solutions. They differ in how these fractions are converted into a final selection:
\begin{itemize}
    \item \textbf{Randomized Rounding} samples seeds according to the calculated fractional values, introducing randomness into the selection.
    \item \textbf{Standard Rounding} deterministically rounds fractional values into an integral selection.
\end{itemize}

We replaced \sys's core algorithm with each method, yielding two AFL++-based variants: \aflpprr (Randomized Rounding) and \aflppsr (Standard Rounding). For implementation, we integrated \texttt{HiGHS} \cite{highs}, a high-performance open-source linear optimization suite, as the LP solver due to its strong performance and convenient C API. We evaluated these two variants on the FuzzBench benchmark by measuring code coverage and runtime overhead over 10 independent 24-hour fuzzing campaigns. To ensure a fair comparison, we invoke the LP-based subset selection at the same frequency as \sys's subset construction during seed scheduling.

\autoref{tab:rr_sr_coverage} presents the mean edge coverage achieved by \aflpprr and \aflppsr, alongside the relative improvement yielded by \sys on each target. It is evident that \sys outperforms both variants on the vast majority of targets, achieving an average coverage gain of 6.97\% over \aflpprr and 5.02\% over \aflppsr. \autoref{tab:rr_sr_overhead} further details the runtime overhead incurred by these approaches during the fuzzing process. \aflpprr and \aflppsr exhibit substantial average overheads of 61.20\% and 58.52\%, respectively, indicating that more than half of the total fuzzing budget is consumed by the algorithm execution itself. In sharp contrast, \sys introduces a negligible overhead of only 1.17\%. Notably, for high-throughput targets such as \texttt{curl}, \texttt{harfbuzz}, and \texttt{libxml}, the two variants consume approximately 90\% of the runtime. This excessive overhead directly translates into a significant performance gap: \sys outperforms \aflpprr in coverage by 11.61\%, 13.79\%, and 26.03\%, and \aflppsr by 6.97\%, 14.43\%, and 15.74\%, respectively. The performance gap is primarily attributed to the need to solve LP relaxations at the frequency required by seed scheduling, which dominates the runtime overhead. Furthermore, in their plain form, these approximation algorithms may fail to cover all frontier nodes during subset selection. Although adding a deterministic repair step can ensure full coverage, it would further increase the overhead. In contrast, \sys avoids expensive LP solving via a lightweight design while ensuring complete frontier coverage. Consequently, \sys proves to be a significantly more efficient and robust solution than these classic approximation algorithms for corpus reduction during fuzzing seed scheduling.

}

\added{
\begin{longfbox}
\textbf{Result 8:} 
On average, \sys achieves 6.97\% higher coverage than \aflpprr and 5.02\% higher coverage than \aflppsr. Moreover, these two variants incur 61.2\% and 58.52\% runtime overhead, respectively.
\end{longfbox}
}

\section{Discussion}
In this section, we analyze the limitations of \sys from both theoretical and implementation perspectives, and identify promising directions for future improvements.

\noindent \textbf{Extensible Scheduling.} Our current approach implements a simple prioritization strategy to order seeds within the reduced set. However, corpus reduction can be naturally integrated with traditional seed scheduling algorithms that incorporate various metrics such as security-critical properties. Such integration could potentially enhance exploration effectiveness by combining the benefits of a minimal yet complete feature set with sophisticated scheduling strategies.

\noindent \textbf{Feature Choice.} While we chose frontier nodes as our feature representation due to their strong correlation with potential coverage gains, this is just one possible feature choice. The current approach requires bitmap-based validation to determine frontier nodes, which introduces additional overhead. Alternative feature representations like basic blocks or edge coverage could potentially offer lower time and space complexity while still effectively guiding corpus reduction.


\section{Related Work}

\noindent\textbf{Set cover problem} 
The Set Cover Problem\cite{caprara2000algorithms,chvatal1979greedy,feige1998threshold} is a classical question in computer science and combinatorial optimization. It involves a universe of elements and a collection of subsets whose union covers the entire universe. The goal is to identify the smallest possible number of these subsets that together contain all elements in the universe. Caprara et al. \cite{caprara2000algorithms} review heuristic and exact algorithms for the Set Covering Problem (SCP), highlighting their key characteristics and comparing their performance on standard benchmark instances. Munagala et al. \cite{munagala2005pipelined} introduce the Pipelined Set Cover problem, a generalization of set cover motivated by query optimization with correlated selections. Among the various algorithmic approaches, randomized rounding techniques for linear programming relaxations have been successfully applied to achieve O(log n) approximation ratios for the set cover problem, matching the best-known polynomial-time approximation bounds while providing probabilistic guarantees on solution quality \cite{vazirani2001approximation}. While the set cover problem has been extensively studied in theoretical research, its application in software testing remains relatively limited. MinSet \cite{rebert2014optimizing} leverages the set cover algorithm solely for selecting initial fuzzing seeds. Similarly, RULF \cite{jiang2021rulf} applies set cover in the context of Rust library fuzzing to merge the API sequence sets, followed by refinement. However, none of these studies explore the use of set cover for seed scheduling during the fuzzing process.


\noindent\textbf{Seed Scheduling} 
Seed scheduling primarily focuses on input prioritization~\cite{Wang2020NotAC, sensitive, wang2021reinforcement}, which aims to rank and select seeds based on various metrics to maximize fuzzing effectiveness. Our approach of seed reduction is orthogonal to these prioritization-based methods, as we focus on reducing the corpus size while existing techniques operate on ranking seeds within the given corpus.

Previous work has explored a range of strategies for input prioritization. Inputs have been prioritized based on edge or path coverage~\cite{lemieux2018fairfuzz, aflfast, entropic, ecofuzz}, execution time~\cite{slowfuzz, perffuzz}, exploitability~\cite{woo13}, memory access patterns~\cite{memfuzz, memlock, Wang2020NotAC}, or combinations of these metrics~\cite{sensitive, wang2021reinforcement}. \texttt{SAVIOR}~\cite{savior} introduces a bug-driven hybrid testing approach that estimates the number of reachable bugs by assuming that all edges are equally likely to be reachable and feasible, regardless of their distance from the seed’s execution path. This assumption does not always hold in real-world programs.

Another line of work prioritizes seeds based on call graphs~\cite{cerebro}. AFLGo~\cite{aflgo} leverages the entire inter-procedural control-flow graph (CFG) to compute the distance between a seed and predefined target locations, which is then used to guide directed fuzzing and allocate mutation budgets. K-Scheduler~\cite{She2022EffectiveSS} also employs an inter-procedural CFG to approximate the number of reachable and feasible edges from a given seed, and utilizes this estimate to inform coverage-guided fuzzing. It introduces the concept of frontier nodes and uses this abstraction to guide seed selection. However, it  suffers from the issue of seed explosion. FOX~\cite{she2024fox} uses frontier-node but focuses on designing an efficient solver to optimize seed mutation, which does not consider corpus reduction to reduce redundancy.

\noindent\textbf{Corpus Reduction} Existing corpus reduction techniques like AFL-Cmin\cite{aflcmin} and MINSET \cite{rebert2014optimizing} minimize test cases while preserving code coverage, primarily for initial seed selection before fuzzing begins. Recent work by Schiller et al. \cite{schiller2025novelty} applies corpus reduction during fuzzer restarts to mitigate input shadowing and discover new initial states. However, these approaches are limited to one-time or restart-point operations. Our work enables continuous corpus reduction during seed scheduling with low overhead and preserved diversity, complementing existing techniques that focus on initial selection or restart scenarios.

\section{Conclusion}
In conclusion, this paper presents a novel corpus reduction technique to address the seed explosion problem in coverage-guided fuzzing. Our solution \sys reduces the corpus to a minimal subset of seeds that covers all features of the original corpus. Through comprehensive evaluation on three popular fuzzers across multiple benchmarks, we demonstrate that \sys effectively mitigates seed explosion and improves fuzzing performance with minimal overhead.

\section*{Data-Availability Statement}
The source code of RandSet is publicly available at \url{https://github.com/Crepuscule-v/RandSet}.

\begin{acks}
We sincerely appreciate the anonymous reviewers for their valuable feedback and guidance. 
The HKUST authors were supported in part by a RGC GRF grant under the contract 16214723.
\end{acks}

\newpage
\appendix

\newpage
\section{Appendix}

\subsection{\added{AFL++ Integration Experiment Results}}

\begin{figure*}[!hbtp]
    \centering
    \includegraphics[width=0.85\linewidth]{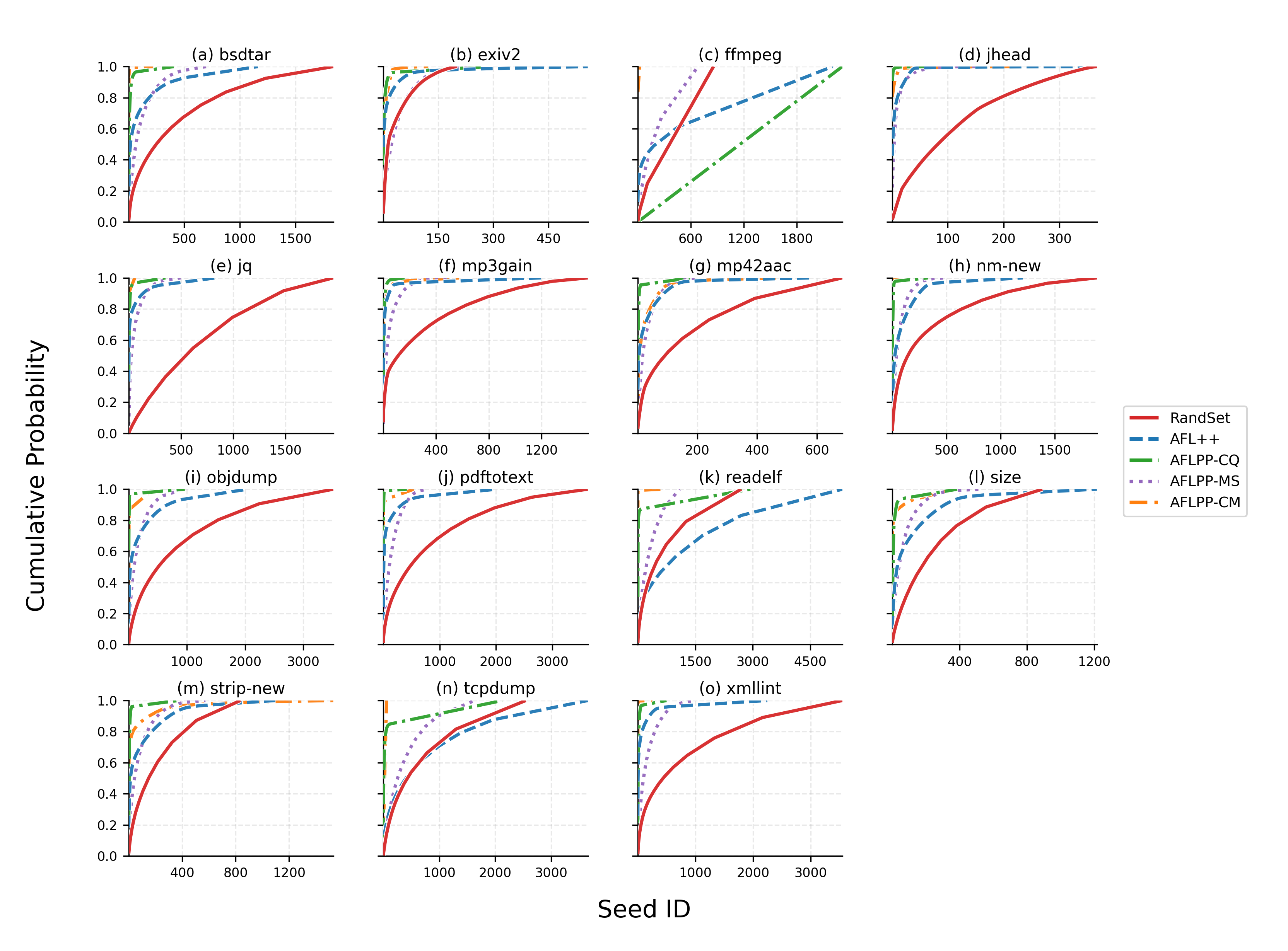}
    \caption{\small \textbf{Cumulative frequency distributions of seeds for \sys against other four baselines on standalone programs over \replaced{10 fuzzing campaigns of 24 hours each}{a 24-hour fuzzing campaign}. The x-axis represents the seed ID sorted in descending order by frequency\replaced{;}{,} the y-axis indicates the cumulative probability of seed selection.}}
    \label{fig:standalone_diversity}
\end{figure*}

\begin{table}[!htbp]
\scriptsize
    \centering
    \caption{\added{\small\textbf{Cumulative number of unique bugs triggered by \sys against all four baselines in FuzzBench programs and standalone programs for 24 hours over 10 runs.}}}
    {
    \begin{tabular}{lccccc}
    \toprule
    \textbf{Targets} & \textbf{AFL++} & \textbf{\aflppminset} & \textbf{\aflppcq} & \textbf{\aflppcmin} & \textbf{\sys} \\
    \midrule
    jhead       & 3   & 3   & 3   & 1   & 3   \\
    jq          & 1   & 1   & 1   & 1   & 1   \\
    mp3gain     & 11  & 9   & 10  & 1   & 11  \\
    pdftotext   & 1   & 1   & 1   & 1   & 1   \\
    tcpdump     & 2   & 1   & 1   & 0   & 2   \\
    bloaty      & 1   & 1   & 1   & 1   & 1   \\
    woff2       & 1   & 0   & 1   & 0   & 2   \\
    \midrule
    \textbf{Total} & \textbf{20} & \textbf{16} & \textbf{18} & \textbf{5} & \textbf{21} \\
    \bottomrule
    \end{tabular}
    }
    \footnotesize
    \label{tab:realwrold_bugs}
\end{table}

\begin{table}[!ht]
\small
    \centering
    \caption{\small\textbf{Total unique seeds for RandSet and four AFL++-based baselines on 15 standalone programs over \replaced{10 fuzzing campaigns of 24 hours each}{a 24-hour fuzzing campaign}. The highest value for each target is in bold.}}
    \begin{tabular}{lrrrrr}
    \toprule
    \textbf{Target} & \textbf{AFL++} & \textbf{AFLPP-MS} & \textbf{AFLPP-CQ} & \textbf{AFLPP-CM} & \textbf{RandSet} \\
    \midrule
    bsdtar      & 1163 & 698  & 406  & 219  & \textbf{1841} \\
    exiv2       & \textbf{557}  & 285  & 289  & 123  & 202  \\
    ffmpeg      & 2208 & 682  & \textbf{2318} & 21  & 858  \\
    jhead       & 344  & 153  & 63   & 225    & \textbf{367}  \\
    jq          & 833  & 497  & 352  & 58  & \textbf{1958} \\
    mp3gain     & 1199 & 514  & 162  & 589  & \textbf{1550} \\
    mp42aac     & 586  & 214  & 175  & 480   & \textbf{685}  \\
    nm-new      & 1206 & 484  & 329  & 173  & \textbf{1888} \\
    objdump     & 2042 & 947  & 963  & 419  & \textbf{3515} \\
    pdftotext   & 2018 & 769  & 418  & 541  & \textbf{3629} \\
    readelf     & \textbf{5335} & 1086 & 2935 & 652  & 2697  \\
    size        & \textbf{1216} & 531  & 386  & 367  & 892  \\
    strip-new   & 1096 & 574  & 372  & \textbf{1530}  & 836  \\
    tcpdump     & \textbf{3654} & 1630 & 2129 & 59 & 2546  \\
    xmllint     & 2250 & 999  & 498  & 257  & \textbf{3552} \\
    \bottomrule
    \end{tabular}
    \label{tab:unique_seed_standalone}
\end{table}

\begin{table}[!ht]
\small
    \setlength{\tabcolsep}{2pt} 
    \centering
    \caption{\small\textbf{ Mean edge coverage of RandSet against AFL++-based baselines on \numstandalone standalone programs for 24 hours over 10 runs. \replaced{We mark the highest number in bold.}{Results are presented as Mean ± Std, with the highest numbers marked in bold.}} }
    
    \begin{tabular}{l *{5}{r @{\,{\scriptsize$\pm$}\,} >{\scriptsize}l}}
    \toprule
    \textbf{Targets} & 
    \multicolumn{2}{c}{\textbf{AFL++}} & 
    \multicolumn{2}{c}{\textbf{\aflppminset}} & 
    \multicolumn{2}{c}{\textbf{\aflppcq}} & 
    \multicolumn{2}{c}{\textbf{\aflppcmin}} & 
    \multicolumn{2}{c}{\textbf{\sys}} \\
    \midrule
    bsdtar      & 4646.2 & 84   & 4083.6 & 805  & 2989.6 & 469  & 2705.2 & 312  & \textbf{4851.2} & 628  \\
    exiv2       & 2801   & 1736 & 4656.2 & 2358 & 4548.8 & 186  & 2969   & 1977 & \textbf{6294}   & 1623 \\
    ffmpeg      & 38845  & 1632 & 35237  & 2083 & \textbf{43458}  & 2624 & 24977  & 5748 & 39347.2 & 3502 \\
    jhead       & 361    & 0    & 361    & 0    & 361    & 0    & 355    & 0    & 361     & 0    \\
    jq          & 3571   & 60   & 3560.4 & 18   & 3531.2 & 56   & 3314   & 134  & \textbf{3576.2} & 12   \\
    mp3gain     & 1560   & 14   & 1564.8 & 13   & 1524.2 & 15   & 1551   & 106  & \textbf{1575.4} & 18   \\
    mp42aac     & 2038   & 32   & 1770.4 & 48   & 1667   & 103  & 1772.6 & 60   & \textbf{2051.6} & 31   \\
    nm-new      & 2621.4 & 20   & 3149.2 & 710  & 2408.4 & 1064 & 1996.5 & 460  & \textbf{3438}   & 609  \\
    objdump     & 6051.4 & 151  & 6247   & 189  & 5750.4 & 423  & 3688   & 784  & \textbf{6288}   & 575  \\
    pdftotext   & 6869   & 35   & 7015.2 & 72   & 6515.6 & 134  & 6449.5 & 114  & \textbf{7290}   & 36   \\
    readelf     & \textbf{6171}   & 97   & 5436   & 344  & 6040   & 254  & 4960.3 & 755  & 6152.2  & 221  \\
    size        & 2443.2 & 155  & \textbf{3007.6} & 120  & 2603   & 100  & 2463   & 292  & 2739.6  & 143  \\
    strip-new   & 2853   & 136  & 3342   & 196  & 2992   & 248  & \textbf{5069.8} & 732  & 3236.4  & 199  \\
    tcpdump     & 7463.4 & 249  & 8111.2 & 456  & 6911.8 & 1087 & 2302   & 1909 & \textbf{8412.2} & 767  \\
    xmllint     & 5331.2 & 65   & 5186.2 & 156  & 5120   & 125  & 4516.2 & 98   & \textbf{7317}   & 982  \\
    \midrule
    \textbf{Mean Gain} & 
    \multicolumn{2}{c}{\textbf{16.58\%}} & 
    \multicolumn{2}{c}{\textbf{9.49\%}} & 
    \multicolumn{2}{c}{\textbf{17.52\%}} & 
    \multicolumn{2}{c}{\textbf{50.54\%}} & 
    \multicolumn{2}{c}{—} \\
    \bottomrule
    \end{tabular}
    \label{tab:standalone}
\end{table}

\begin{table}[!ht]
\small
    \centering
    \caption{\small\textbf{Runtime overhead  (\%) of \sys compared to three baselines on standalone targets. Mean Ratio is the average overhead across all targets.}}
    \begin{tabular}{lrrrr}
    \toprule
    \textbf{Targets} & \textbf{\aflppminset} & \textbf{\aflppcq} & \textbf{\aflppcmin} & \textbf{\sys} \\
    \midrule
    bsdtar      & 4.91   & 0.02   & 17.63 & 2.10 \\
    exiv2       & 5.02   & 0.01   & 12.90 & 0.21 \\
    ffmpeg      & 67.34  & 0.33   & 10.33 & 21.70 \\
    jhead       & 0.20   & 0.00   & 14.72 & 0.34 \\
    jq          & 5.89   & 0.00   & 14.55 & 2.16 \\
    mp3gain     & 3.47   & 0.00   & 14.57 & 1.54 \\
    mp42aac     & 0.66   & 0.01   & 0.14  & 0.20 \\
    nm-new      & 5.52   & 0.02   & 17.47 & 2.36 \\
    objdump     & 44.56  & 0.17   & 14.96 & 8.25 \\
    pdftotext   & 50.45  & 0.10   & 16.65 & 0.32 \\
    readelf     & 66.26  & 0.26   & 22.40 & 3.86 \\
    size        & 5.83   & 0.01   & 19.16 & 0.18 \\
    strip-new   & 5.33   & 0.02   & 18.38 & 0.33 \\
    tcpdump     & 35.50  & 0.07   & 10.28 & 1.66 \\
    xmllint     & 32.55  & 0.03   & 19.79 & 13.75 \\
    \midrule
    \textbf{Mean Ratio} & \textbf{22.23}  & \textbf{0.07}   & \textbf{14.93} & \textbf{3.93} \\
    \bottomrule
    \end{tabular}
    \label{tab:standalone_overhead}
\end{table}

\begin{table}[t]
\centering
\footnotesize
\caption{\textbf{\added{Runtime overhead (\%) of frontier node calculation for AFL++ integrations on both the FuzzBench and standalone targets, measured as the percentage of total fuzzing time.}}}
\label{tab:aflpp_frontier_overhead}
\renewcommand{\arraystretch}{1.08}
\setlength{\tabcolsep}{6pt}

\begin{tabular}{lr|lr|lr|lr}
\toprule
\textbf{Targets} & \textbf{Overhead} &
\textbf{Targets} & \textbf{Overhead} &
\textbf{Targets} & \textbf{Overhead} &
\textbf{Targets} & \textbf{Overhead} \\
\midrule

\multicolumn{2}{c|}{\textbf{FuzzBench Targets}} & re2      & 0.0007 & woff2      & 0.0002 & sqlite3    & 0.0018 \\ \cline{1-2}
bloaty    & 0.0008 & curl       &  0.0010      & stbi       & 0.0001 & harfbuzz   & 0.0061 \\
jsoncpp   & 0.0001 & lcms       & 0.0001 & libjpeg    & 0.0002 & libpcap    & 0.0001 \\
libpng    & 0.0001 & libxml    & 0.0041 & libxslt    & 0.0010 & zlib       & 0.0001 \\
openh264  & 0.0032 & openssl    & 0.0005 & openthread & 0.0002 & vorbis     & 0.0002 \\
\midrule

\multicolumn{2}{c|}{\textbf{Standalone Targets}} & tcpdump  & 0.0009 & mp3gain    & 0.0001 & jhead      & 0.0001 \\ \cline{1-2}
mp42aac   & 0.0001 & xmllint    & 0.0023 & bsdtar     & 0.0009 & exiv2      & 0.0009 \\
ffmpeg    & 0.0264 & jq         & 0.0001 & nm-new     & 0.0010 & objdump    & 0.0015 \\
pdftotext & 0.0017 & readelf    & 0.0023 & size       & 0.0004 & strip-new  & 0.0008 \\
\bottomrule
\end{tabular}
\end{table}

\begin{table}[]
\centering
\footnotesize
\caption{\textbf{\added{CFG construction time (s) of \sys for AFL++ integrations on both the FuzzBench and standalone targets.}}}
\label{tab:aflpp_cfg_time}
\renewcommand{\arraystretch}{1.1}
\begin{tabular}{lr|lr|lr|lr}
\toprule
\textbf{Targets} & \textbf{Time } & \textbf{Targets} & \textbf{Time } & \textbf{Targets} & \textbf{Time } & \textbf{Targets} & \textbf{Time } \\
\midrule
\multicolumn{2}{c|}{\textbf{FuzzBench Targets}} & re2             & 4.28  & woff2          & 9.81  & sqlite3        & 40.22 \\ \cline{1-2}
bloaty        & 77.32  & curl & 63.51 & stbi           & 0.02  & harfbuzz       & 26.42 \\
jsoncpp       & 4.11   & lcms            & 3.88  & libjpeg        & 8.19  & libpcap        & 5.60  \\
libpng        & 0.02   & libxml        & 41.04 & libxslt        & 34.46 & zlib           & 0.01  \\
openh264     & 8.10   & openssl        & 33.47 & openthread     & 26.02 & vorbis         & 5.09  \\
\midrule
\multicolumn{2}{c|}{\textbf{Standalone Targets}} & tcpdump         & 12.10 & mp3gain        & 0.15  & jhead          & 0.95  \\ \cline{1-2}
mp42aac       & 8.06   & xmllint         & 46.28 & bsdtar         & 14.25 & exiv2          & 56.96 \\
ffmpeg        & 257.48 & jq              & 5.07  & nm-new         & 22.91 & objdump        & 42.90 \\
pdftotext     & $<$0.01& readelf         & 18.28 & size           & 26.43 & strip-new      & 30.06 \\
\bottomrule
\end{tabular}
\end{table}

\begin{figure*}[!]
    \centering
    \includegraphics[width=0.85\linewidth]{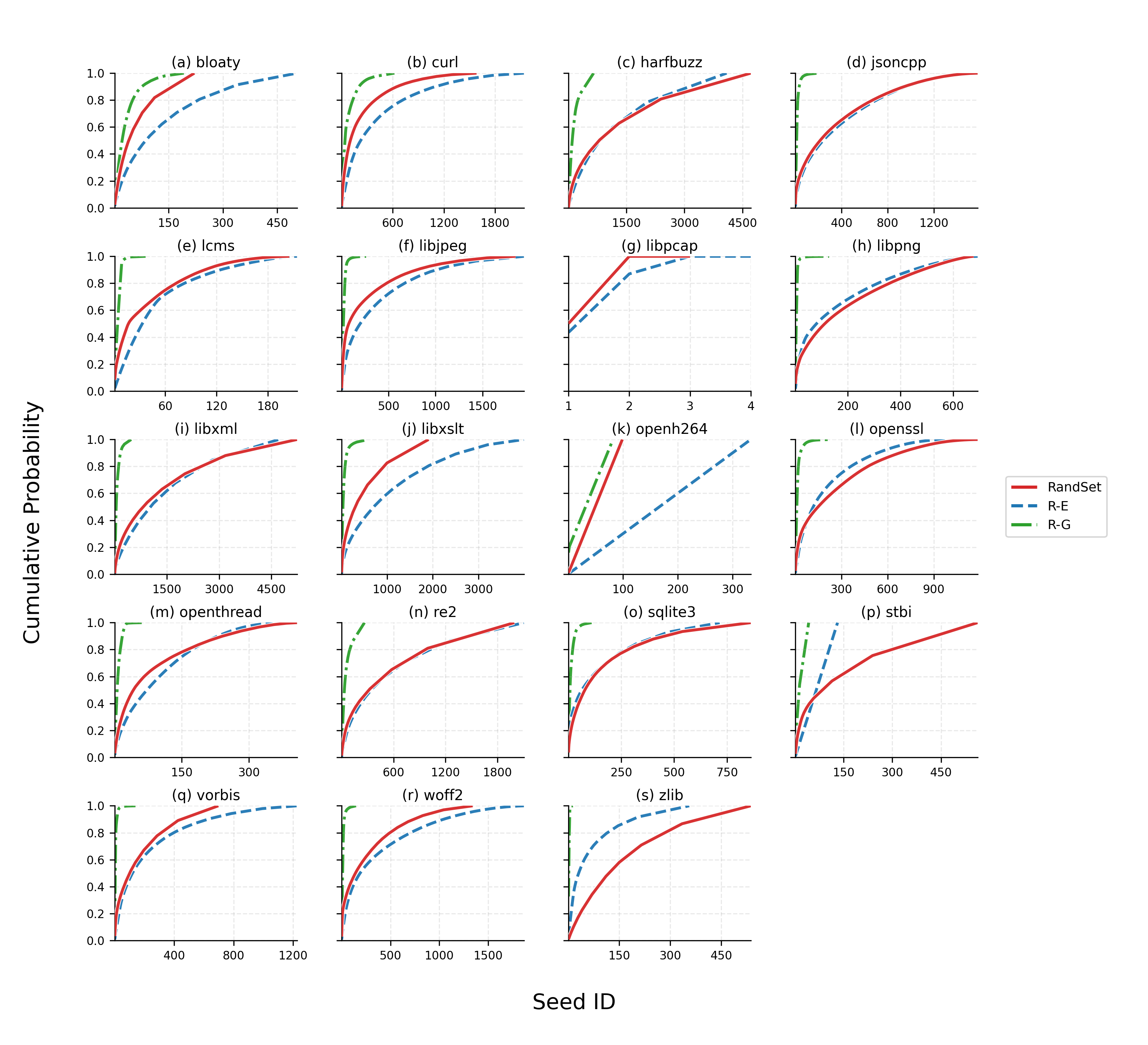}
    \caption{\small \textbf{Cumulative frequency distributions of seeds for \sys against other three variants on FuzzBench programs over \replaced{10 separate 1-hour fuzzing campaigns.}{a 1-hour fuzzing campaign}. The x-axis represents the seed ID sorted in descending order by frequency\replaced{;}{,} the y-axis indicates the cumulative probability of seed selection.}}
    \label{fig:ablatation_diversity}
\end{figure*}

\clearpage
\subsection{\added{LibAFL and Centipede Integration Experiment Results }}

\begin{table}[!htbp]
\scriptsize
\caption{\small\textbf{Mean edge coverage of \sys against Non-AFL++ baseline fuzzers (LibAFL and Centipede) on FuzzBench programs for 24 hours over 10 runs. \added{Results are presented as Mean $\pm$ Std.} We mark the highest mean coverage in bold.}}
    \centering
    \setlength{\tabcolsep}{1.5pt}
    
    \begin{tabular}{
        l r@{\,{\scriptsize$\pm$}\,}>{\scriptsize}l r@{\,{\scriptsize$\pm$}\,}>{\scriptsize}l ||
        l r@{\,{\scriptsize$\pm$}\,}>{\scriptsize}l r@{\,{\scriptsize$\pm$}\,}>{\scriptsize}l ||
        l r@{\,{\scriptsize$\pm$}\,}>{\scriptsize}l r@{\,{\scriptsize$\pm$}\,}>{\scriptsize}l
    }
    \toprule
        \multicolumn{15}{c}{\textbf{\libafl Based}} \\
        \midrule
        \textbf{Targets} & \multicolumn{2}{c}{\textbf{Baseline}} & \multicolumn{2}{c||}{\textbf{\sys}} & 
        \textbf{Targets} & \multicolumn{2}{c}{\textbf{Baseline}} & \multicolumn{2}{c||}{\textbf{\sys}} & 
        \textbf{Targets} & \multicolumn{2}{c}{\textbf{Baseline}} & \multicolumn{2}{c}{\textbf{\sys}} \\ 
        \midrule
        bloaty   & 4,466 & 31 & \textbf{4,608} & 93       & openthread & 4,315 & 19 & \textbf{4,435} & 14       & libxslt & \textbf{4,624} & 251 & 4,573 & 231 \\
        curl     & \multicolumn{2}{c}{/} & \multicolumn{2}{c||}{/} & re2        & 6,457 & 76 & \textbf{6,463} & 57       & sqlite3 & 11,977 & 739 & \textbf{12,064} & 577 \\
        harfbuzz & 19,033 & 298 & \textbf{19,480} & 317 & zlib       & 713 & 13   & \textbf{716} & 11         & stbi    & 1,957 & 12  & \textbf{2,224} & 25 \\
        jsoncpp  & 1,228 & 18 & \textbf{1,235} & 11       & libpcap    & \textbf{52} & 0 & \textbf{52} & 3       & vorbis  & \textbf{2,165} & 47 & 2,160 & 73 \\
        lcms     & 1,191 & 14 & \textbf{1,294} & 41       & libjpeg    & \textbf{6,059} & 83 & 5,830 & 112      & woff2   & \textbf{2,378} & 22 & 2,364 & 10 \\
        \cline{11-15} 
        openh264 & \multicolumn{2}{c}{/} & \multicolumn{2}{c||}{/} & libpng     & \textbf{1,253} & 31 & \textbf{1,253} & 28 & \multicolumn{1}{l}{\multirow{2}{*}{\textbf{Mean gain}}} & \multicolumn{2}{c}{\multirow{2}{*}{\textbf{\maxlibaflfuzzbenchcov}}} & \multicolumn{2}{c}{\multirow{2}{*}{-}} \\
        openssl  & \multicolumn{2}{c}{/} & \multicolumn{2}{c||}{/} & libxml     & 8,182 & 453 & \textbf{10,756} & 593   & \multicolumn{5}{l}{} \\
        \midrule\midrule
        
        \multicolumn{15}{c}{\textbf{\centipede Based}}\\
        \midrule
        \textbf{Targets} & \multicolumn{2}{c}{\textbf{Baseline}} & \multicolumn{2}{c||}{\textbf{\sys}} & 
        \textbf{Targets} & \multicolumn{2}{c}{\textbf{Baseline}} & \multicolumn{2}{c||}{\textbf{\sys}} & 
        \textbf{Targets} & \multicolumn{2}{c}{\textbf{Baseline}} & \multicolumn{2}{c}{\textbf{\sys}} \\ 
        \midrule
        bloaty   & 2,876 & 317 & \textbf{3,456} & 221 & openthread & 4,546 & 302 & \textbf{5,224} & 466 & libxslt & 4,401 & 185 & \textbf{4,803} & 172 \\
        curl     & \multicolumn{2}{c}{/} & \multicolumn{2}{c||}{\textbf{/}}       & re2        & 5,429 & 69  & \textbf{5,549} & 83  & sqlite3 & \textbf{54} & 0 & \textbf{54} & 0 \\
        harfbuzz & 18,555 & 844 & \textbf{19,778} & 512 & zlib     & 693 & 6   & \textbf{702} & 4   & stbi    & 1,636 & 92  & \textbf{1,999} & 232 \\
        jsoncpp  & \textbf{33} & 0 & \textbf{33} & 0  & libpcap    & 1,773 & 1213 & \textbf{2,080} & 234 & vorbis  & 1,960 & 54  & \textbf{1,988} & 49 \\
        lcms     & 809 & 62  & \textbf{1,045} & 236     & libjpeg    & 2,414 & 139 & \textbf{2,711} & 82  & woff2   & 1,985 & 54  & \textbf{2,000} & 60 \\
        \cline{11-15} 
        openh264 & \textbf{11,214} & 57 & 11,042 & 54 & libpng     & \textbf{100} & 0 & \textbf{100} & 0 & \multicolumn{1}{l}{\multirow{2}{*}{\textbf{Mean gain}}} & \multicolumn{2}{c}{\multirow{2}{*}{\textbf{\maxcentipedefuzzbenchcov}}} & \multicolumn{2}{c}{\multirow{2}{*}{-}} \\
        openssl  & 3,710 & 25 & \textbf{3,771} & 52     & libxml     & 8,857 & 457 & \textbf{8,969} & 331 & \multicolumn{5}{l}{} \\
        \bottomrule
    \end{tabular}
    \label{eval:fuzzbench_nonaflpp}
\end{table}

\begin{table}[!htbp]
\scriptsize
\caption{\small\textbf{Corpus reduction effectiveness of \sys on Fuzzbench targets across other Non-AFL++ fuzzers after 24-hour fuzzing campaigns. For each target, Total and Subset show the average final corpus size and dynamic set size (rounded to integers) across 10 runs, while Ratio represents the percentage of retained seeds calculated from raw data. The results are grouped by base fuzzer (\libafl and \centipede).}}
    \centering
    \setlength{\tabcolsep}{1.5pt}
    \begin{tabular}{lrrr||lrrr||lrrr}
    \toprule
    \multicolumn{12}{c}{\textbf{\libafl Based}} \\
    \midrule
    \textbf{Targets} & \textbf{Total} & \textbf{Subset} & \textbf{Ratio(\%)} & \textbf{Targets} & \textbf{Total} & \textbf{Subset} & \textbf{Ratio(\%)} & \textbf{Targets} & \textbf{Total} & \textbf{Subset} & \textbf{Ratio(\%)} \\ 
    \midrule
    woff2 & 1,211 & 142 & 11.76 & libxml & 7,697 & 682 & 8.86 & curl & / & / & / \\
    harfbuzz & 7,585 & 1,202 & 15.84 & jsoncpp & 1,221 & 99 & 8.14 & stbi & 782 & 202 & 25.85 \\
    openssl & / & / & / & vorbis & 797 & 150 & 18.84 & lcms & 164 & 45 & 27.53 \\
    bloaty & 806 & 168 & 20.85 & libpcap & 7 & 3 & 48.57 & openh264 & / & / & / \\
    openthread & 664 & 84 & 12.68 & zlib & 462 & 76 & 16.41 & libjpeg & 2,283 & 359 & 15.73 \\
    \cline{9-12} 
    libxslt & 3,932 & 40 & 1.02 & re2 & 2,715 & 332 & 12.22 & \multicolumn{2}{l}{\multirow{2}{*}{\textbf{Mean Ratio}}} & \multicolumn{2}{r}{\multirow{2}{*}{\textbf{\libaflfuzzbenchdistill}}} \\
    libpng & 428 & 70 & 16.36 & sqlite3 & 1,085 & 169 & 15.58 & \multicolumn{2}{l}{} & \multicolumn{2}{r}{} \\
    \midrule\midrule
    
    \multicolumn{12}{c}{\textbf{\centipede Based}} \\
    \midrule
    \textbf{Targets} & \textbf{Total} & \textbf{Subset} & \textbf{Ratio(\%)} & \textbf{Targets} & \textbf{Total} & \textbf{Subset} & \textbf{Ratio(\%)} & \textbf{Targets} & \textbf{Total} & \textbf{Subset} & \textbf{Ratio(\%)} \\ 
    \midrule
    woff2 & 7,013 & 132 & 1.88 & libxml & 27,130 & 570 & 2.10 & curl & / & / & / \\
    harfbuzz & 55,054 & 1429 & 2.60 & jsoncpp & 52 & 2 & 3.88 & stbi & 5,932 & 151 & 2.54 \\
    openssl & 5,845 & 151 & 2.58 & vorbis & 5,674 & 115 & 2.03 & lcms & 2,210 & 41 & 1.84 \\
    bloaty & 6,750 & 73 & 1.08 & libpcap & 3,742 & 148 & 3.96 & openh264 & 21,780 & 522 & 2.40 \\
    openthread & 5,534 & 131 & 2.37 & zlib & 3,386 & 69 & 2.03 & libjpeg & 7,888 & 163 & 2.06 \\
    \cline{9-12} 
    libxslt & 10721 & 264 & 2.46 & re2 & 13,346 & 222 & 1.66 & \multicolumn{2}{l}{\multirow{2}{*}{\textbf{Mean Ratio}}} & \multicolumn{2}{r}{\multirow{2}{*}{\textbf{\centipedefuzzbenchdistill}}} \\
    libpng & 342 & 7 & 1.99 & sqlite3 & 188 & 4 & 2.02 & \multicolumn{2}{l}{} & \multicolumn{2}{r}{} \\
    \bottomrule
    \end{tabular}
    \label{tab:fuzzbench_reduction_non_aflpp_fuzzer}
\end{table}

\begin{table}[!htbp]
\small
\caption{\small\textbf{Runtime overhead (\%) of \sys against other Non-AFL++ fuzzers on FuzzBench programs, measured as the proportion of corpus reduction in total fuzzing time. The results are grouped by base fuzzer (LibAFL and Centipede).}}
    \centering
    \resizebox{\columnwidth}{!}{
    \setlength{\tabcolsep}{1.5pt}
    \begin{tabular}{lr|lr|lr|lr|lr}
    \toprule
        \textbf{Targets} & \textbf{Overhead} & \textbf{Targets} & \textbf{Overhead} & \textbf{Targets} & \textbf{Overhead} & \textbf{Targets} & \textbf{Overhead} & \textbf{Targets} & \textbf{Overhead} \\ 
        \midrule
        \multicolumn{10}{c}{\textbf{\libafl Based}} \\
        \midrule
        bloaty & 1.310 & lcms & 1.140 & libjpeg & 10.920 & re2 & 12.960 & stbi & 0.120 \\
        curl & / & openthread & 1.760 & libpng & 8.770 & libxml & 11.390 & vorbis & 2.010 \\
        harfbuzz & 16.230 & zlib & 24.040 & openh264 & / & libxslt & 21.090 & woff2 & 19.740 \\
        \cline{9-10} 
        jsoncpp & 10.940 & libpcap & 2.020 & openssl & / & sqlite3 & 2.740 & \textbf{Average} & \textbf{\libaflfuzzbenchover} \\
        \midrule\midrule
        
        \multicolumn{10}{c}{\textbf{\centipede Based}}\\
        \midrule
        \textbf{Targets} & \textbf{Overhead} & \textbf{Targets} & \textbf{Overhead} & \textbf{Targets} & \textbf{Overhead} & \textbf{Targets} & \textbf{Overhead} & \textbf{Targets} & \textbf{Overhead} \\ 
        \midrule
        bloaty & 0.072 & lcms & 0.019 & libjpeg & 0.109 & re2 & 0.044 & stbi & 0.002 \\
        curl & / & openthread & 0.179 & libpng & 0.033 & libxml & 1.184 & vorbis & 0.051 \\
        harfbuzz & 2.994 & zlib & 0.188 & openh264 & 0.119 & libxslt & 0.469 & woff2 & 0.125 \\
        \cline{9-10} 
        jsoncpp & 0.021 & libpcap & 0.248 & openssl & 0.454 & sqlite3 & 0.008 & \textbf{Average} & \textbf{\centipedefuzzbenchover} \\
        \bottomrule
    \end{tabular}
    \label{tab:overheadfuzzbench_non_aflpp_fuzzers}
    }
\end{table}

\begin{table}[!htbp]
\small
\caption{\small\textbf{\added{CFG construction time (s) of \sys for LibAFL and Centipede integrations on FuzzBench targets.}}}
    \centering
    \setlength{\tabcolsep}{1.5pt}
    \begin{tabular}{lr|lr|lr|lr|lr}
    \toprule
        \textbf{Targets} & \textbf{Time } & \textbf{Targets} & \textbf{Time } & \textbf{Targets} & \textbf{Time } & \textbf{Targets} & \textbf{Time } & \textbf{Targets} & \textbf{Time } \\ 
        \midrule
        \multicolumn{10}{c}{\textbf{\libafl Based}} \\
        \midrule
        bloaty   & 1.07 & lcms       & 1.16    & libjpeg  & 0.22 & re2     & 0.08 & stbi   & 0.04 \\
        curl     & /    & openthread & 0.23 & libpng   & 0.04 & libxml  & 0.45 & vorbis & 0.05 \\
        harfbuzz & 0.43 & zlib       & 0.01 & openh264 & /    & libxslt & 0.29 & woff2  & 0.10 \\
        jsoncpp  & 0.05 & libpcap    &  0.93    & openssl  & /    & sqlite3 & 0.34 &        &      \\
        \midrule\midrule
        
        \multicolumn{10}{c}{\textbf{\centipede Based}}\\
        \midrule
        \textbf{Targets} & \textbf{Time } & \textbf{Targets} & \textbf{Time } & \textbf{Targets} & \textbf{Time } & \textbf{Targets} & \textbf{Time } & \textbf{Targets} & \textbf{Time } \\ 
        \midrule
        bloaty   & 0.05     & lcms       & $< 0.01$ & libjpeg  & $< 0.01$ & re2     & $< 0.01$ & stbi   & $< 0.01$ \\
        curl     & /        & openthread & 0.02     & libpng   & $< 0.01$ & libxml  & 0.04     & vorbis & $< 0.01$ \\
        harfbuzz & 0.02     & zlib       & $< 0.01$ & openh264 & $< 0.01$ & libxslt & 0.03     & woff2  & $< 0.01$ \\
        jsoncpp  & $< 0.01$ & libpcap    & $< 0.01$ & openssl  & 0.04     & sqlite3 & $< 0.01$ &        &          \\
        \bottomrule
    \end{tabular}
    \label{tab:centipede_libafl_cfg_construction_time}
\end{table}

\begin{table}[!htbp]
\small
\caption{\small\textbf{\added{Runtime overhead (\%) of calculating frontier nodes for LibAFL and Centipede integrations on FuzzBench targets, measured as the percentage of total fuzzing time.}}}
    \centering
    \resizebox{\columnwidth}{!}{
    \setlength{\tabcolsep}{1.5pt}
    \begin{tabular}{lr|lr|lr|lr|lr}
    \toprule
        \textbf{Targets} & \textbf{Overhead} & \textbf{Targets} & \textbf{Overhead} & \textbf{Targets} & \textbf{Overhead} & \textbf{Targets} & \textbf{Overhead} & \textbf{Targets} & \textbf{Overhead} \\ 
        \midrule
        \multicolumn{10}{c}{\textbf{\libafl Based}} \\
        \midrule
        bloaty   & 0.632  & lcms       &   0.242     & libjpeg  & 3.681  & re2     & 5.302  & stbi   & 0.062 \\
        curl     & /       & openthread & 0.65  & libpng   & 1.793  & libxml  & 2.968  & vorbis & 1.304 \\
        harfbuzz & 11.351 & zlib       & 3.894  & openh264 & /       & libxslt & 11.203 & woff2  & 16.802 \\
        \cline{9-10}
        jsoncpp  & 3.522  & libpcap    &  1.972      & openssl  & /       & sqlite3 & 2.004  &   \textbf{Average}     &   4.211     \\
        \midrule\midrule
        
        \multicolumn{10}{c}{\textbf{\centipede Based}}\\
        \midrule
        \textbf{Targets} & \textbf{Overhead} & \textbf{Targets} & \textbf{Overhead} & \textbf{Targets} & \textbf{Overhead} & \textbf{Targets} & \textbf{Overhead} & \textbf{Targets} & \textbf{Overhead} \\ 
        \midrule
        bloaty   & 0.030 & lcms       & 0.015 & libjpeg  & 0.044 & re2     & 0.023 & stbi   & 0.002 \\
        curl     & /     & openthread & 0.148 & libpng   & 0.039 & libxml  & 0.546 & vorbis & 0.029 \\
        harfbuzz & 1.199 & zlib       & 0.207 & openh264 & 0.065 & libxslt & 0.164 & woff2  & 0.084 \\
        \cline{9-10}
        jsoncpp  & 0.006 & libpcap    & 0.134 & openssl  & 0.360 & sqlite3 & 0.005 & \textbf{Average} & 0.172 \\
        \bottomrule
    \end{tabular}
    \label{tab:centipede_libafl_frontier_node_overhead}
    }
\end{table}

\clearpage
\subsection{\added{Ablation Study Experiment Results}}
\begin{table}[!ht]
\small
\centering
\footnotesize
\caption{\small\textbf{Ablation study comparing total unique seeds of different \sys variants on FuzzBench targets \replaced{Results are based on 10 one-hour fuzzing campaigns}{over a one-hour fuzzing campaign}. We mark the highest number in bold.}}
    \begin{tabular}{lrrr}
    \toprule
    \textbf{Targets} & \textbf{\sys} & \textbf{R-E} & \textbf{R-G} \\
    \midrule
    bloaty & 221 & \textbf{505} & 192 \\
    curl & 1582 & \textbf{2137} & 619 \\
    harfbuzz & \textbf{4714} & 4090 & 649 \\
    jsoncpp & \textbf{1579} & 1549 & 200 \\
    lcms & 205 & \textbf{214} & 37 \\
    libjpeg & 1852 & \textbf{1940} & 261 \\
    libpcap & 3 & \textbf{4} & 3 \\
    libpng & 678 & \textbf{693} & 129 \\
    libxml & \textbf{5241} & 4717 & 472 \\
    libxslt & 1908 & \textbf{3999} & 588 \\
    openh264 & 99 & \textbf{333} & 82 \\
    openssl & \textbf{1185} & 975 & 229 \\
    openthread & \textbf{407} & 369 & 61 \\
    re2 & 1993 & \textbf{2104} & 264 \\
    sqlite3 & \textbf{862} & 715 & 110 \\
    stbi & \textbf{563} & 131 & 42 \\
    vorbis & 697 & \textbf{1226} & 140 \\
    woff2 & 1341 & \textbf{1865} & 148 \\
    zlib & \textbf{537} & 356 & 14 \\
    \bottomrule
    \end{tabular}
\label{tab:ablatoin_total_unique_seeds}
\end{table}

\begin{table}[!ht]
\small
\centering
\footnotesize
\caption{\small\textbf{Ablation study comparing subset ratio (\%) of different \sys variants on FuzzBench targets. Results are based on 10 one-hour fuzzing campaigns. We mark the smallest number in bold.}}
\begin{tabular}{lrrr}
\toprule
\textbf{Targets} & \textbf{RandSet} & \textbf{R-G} & \textbf{R-E} \\
\midrule
bloaty            & 12.53 & \textbf{6.65} & 23.25 \\
curl              & 7.29  & \textbf{4.93} & 19.41 \\
harfbuzz          & 4.31  & \textbf{2.81} & 17.55 \\
jsoncpp           & 2.99  & \textbf{1.83} & 10.00 \\
lcms              & 5.33  & \textbf{4.96} & 20.67 \\
libjpeg           & 3.30  & \textbf{2.39} & 9.96 \\
libpcap           & \textbf{66.67} & \textbf{66.67} & 100.00 \\
libpng            & 2.85  & \textbf{2.04} & 10.82 \\
libxml            & 2.22  & \textbf{1.28} & 8.28 \\
libxslt           & 2.81  & \textbf{0.97} & 7.52 \\
openh264          & 4.14  & \textbf{2.37} & 17.43 \\
openssl           & 3.28  & \textbf{2.48} & 17.05 \\
openthread        & 4.93  & \textbf{4.12} & 22.77 \\
re2               & 2.79  & \textbf{1.47} & 9.66 \\
sqlite3           & 1.65  & \textbf{0.97} & 7.42 \\
stbi              & 4.08  & \textbf{3.39} & 14.31 \\
vorbis            & 2.47  & \textbf{1.09} & 15.15 \\
woff2             & 2.52  & \textbf{1.91} & 11.90 \\
zlib              & 1.02  & \textbf{0.51} & 8.31 \\
\midrule
\textbf{Mean Ratio}   & 7.22 & \textbf{5.94} & 18.50 \\
\bottomrule
\end{tabular}
\label{tab:fuzzbenchablation_subset_ratio}
\end{table}

\begin{table}[!ht]
\small
\footnotesize
\caption{\small\textbf{Ablation study comparing code coverage of different \sys variants (\sysgreedyshort: greedy set construction, \sysedgefeatureshort: change frontier to edge feature) on FuzzBench targets. Results are based on 10 one-hour fuzzing campaigns. We mark the highest number in bold.}}
\begin{tabular}{lrrr}
\toprule
\textbf{Targets} & \textbf{\sys} & \textbf{\sysgreedyshort} & \textbf{\sysedgefeatureshort} \\ 
\midrule
bloaty      & 2,936.8 & \textbf{2,937.2} & 2,924.2 \\
curl        & \textbf{12,054.8} & 10,890.2 & 11,759.4 \\
harfbuzz    & 19,698.4 & 17,542.6 & \textbf{19,774.2} \\
jsoncpp     & 1,335.6 & \textbf{1,337.4} & 1,318.6 \\
lcms        & \textbf{1,087.6} & 1,050.6 & 903.4 \\
libjpeg     & 2,912.4 & 2,810.2 & \textbf{2,925.6} \\
libpcap     & 43 & 42 & \textbf{44.8} \\
libpng      & 1,351.4 & 1,352.6 & \textbf{1,360.4} \\
libxml      & \textbf{9,130.4} & 8,381.4 & 8,386.2 \\
libxslt     & \textbf{4,390} & 4,274 & 4,302 \\
openh264    & \textbf{13,147.2} & 10,459.2 & 11,908 \\
openssl     & 4,537.8 & \textbf{4,573.2} & 4,524.8 \\
openthread  & 4,678.2 & \textbf{4,698.8} & 4,646 \\
re2         & 6,184.4 & \textbf{6,215.8} & 6,185.4 \\
sqlite3     & \textbf{11,430.8} & 11,412.6 & 11,356.6 \\
stbi        & 1,582.8 & 1,569.2 & \textbf{1,664.2} \\
vorbis      & 2,024.6 & 2,014 & \textbf{2,030.4} \\
woff2       & \textbf{2,330.6} & 2,201 & 2,327.4 \\
zlib        & 877 & \textbf{878.2} & 877.6 \\
\midrule
\textbf{Mean Gain}  & —  & \textbf{3.96\%} & \textbf{1.94\%} \\
\bottomrule
\end{tabular}
\label{tab:fuzzbenchablation_coverage}
\end{table}

\clearpage
\subsection{\added{RQ8 Experiment Results}}
\label{sec:rr_sr_results}
\begin{table}[!htbp]
\centering
\footnotesize
\caption{\textbf{\added{Mean edge coverage of \aflpprr and \aflppsr on FuzzBench programs for 24 hours over 10 runs. The Gain columns indicate the percentage improvement of \sys over the respective variants.}}}
\label{tab:rr_sr_coverage}
\begin{tabular}{lrrrr}
\toprule
\textbf{Targets} & \textbf{\aflpprr} & \textbf{Gain (\%)} & \textbf{\aflppsr} & \textbf{Gain (\%)} \\
\midrule
bloaty       & 3246  & 5.35\%  & 3282  & 4.30\% \\
curl         & 12954 & 11.61\% & 13634 & 6.97\% \\
harfbuzz     & 20728 & 13.79\% & 20576 & 14.43\% \\
jsoncpp      & 1340  & 0.10\%  & 1342  & -0.04\% \\
lcms         & 1126  & 19.82\% & 1440  & -2.53\% \\
libjpeg      & 2840  & 12.27\% & 2958  & 8.63\% \\
libpcap      & 42    & 14.63\% & 42    & 14.63\% \\
libpng       & 1360  & 0.07\%  & 1361  & 0.00\% \\
libxml       & 9708  & 26.03\% & 11058 & 15.74\% \\
libxslt      & 4413  & 4.71\%  & 4476  & 3.35\% \\
openh264     & 13884 & -0.43\% & 13840 & -0.11\% \\
openssl      & 4678  & 1.04\%  & 4656  & 1.51\% \\
openthread   & 4885  & 2.96\%  & 5028  & 0.12\% \\
re2          & 6267  & 0.06\%  & 6245  & 0.41\% \\
sqlite3      & 12065 & 0.43\%  & 11936 & 1.49\% \\
stbi         & 2311  & 17.41\% & 2210  & 21.02\% \\
vorbis       & 2029  & 0.67\%  & 2030  & 0.62\% \\
woff2        & 2407  & 2.09\%  & 2334  & 5.06\% \\
zlib         & 884   & -0.16\% & 884   & -0.16\% \\
\midrule
\textbf{Mean Gain} &       & \textbf{6.97\%} &       & \textbf{5.02\%} \\
\bottomrule
\end{tabular}
\end{table}

\begin{table}[!ht]
\footnotesize
    \centering
    \caption{\small\textbf{\added{Runtime overhead (\%) of \aflpprr and \aflppsr on FuzzBench targets, measured as the percentage of total fuzzing time. Mean Ratio is the average overhead across all targets.}}}
    \begin{tabular}{lrr}
    \toprule
    \textbf{Targets} & \textbf{\textsc{AFL++-RR}} & \textbf{\textsc{AFL++-SR}} \\
    \midrule
    bloaty      & 19.46 & 17.83 \\
    curl        & 90.19 & 90.53 \\
    harfbuzz    & 93.65 & 94.10 \\
    jsoncpp     & 29.98 & 35.51 \\
    lcms        & 48.96 & 33.85 \\
    libjpeg     & 61.76 & 81.38 \\
    libpcap     & 86.39 & 77.63 \\
    libpng      & 88.68 & 83.55 \\
    libxml      & 89.03 & 88.47 \\
    libxslt     & 72.77 & 76.46 \\
    openh264    & 24.27 & 20.74 \\
    openssl     & 87.48 & 86.83 \\
    openthread  & 73.09 & 72.42 \\
    re2         & 72.60 & 52.23 \\
    sqlite3     & 72.07 & 76.03 \\
    stbi        & 11.56 & 7.62  \\
    vorbis      & 25.10 & 27.40 \\
    woff2       & 72.24 & 57.88 \\
    zlib        & 43.53 & 31.33 \\
    \midrule
    \textbf{Mean Ratio} & \textbf{61.20} & \textbf{58.52} \\
    \bottomrule
    \end{tabular}
    \label{tab:rr_sr_overhead}
\end{table}

\clearpage
\subsection{\added{Mann-Whitney U Test Results}}
\label{sec:appendix_mwu}
\added{To validate that the code coverage improvements achieved by \sys are statistically significant rather than due to the inherent randomness of fuzzing, we conducted Mann-Whitney U tests, following the recommendations of Klees et al.~\cite{klees2018evaluating}. The results for AFL++ integrations are detailed in \autoref{tab:fuzzbench_mwu} and \autoref{tab:standalone_mwu} for FuzzBench and standalone programs, respectively, while the results for LibAFL and Centipede integrations are presented in \autoref{tab:libafl_centipede_mwu}.}

\begin{table}[!ht]
\small
    \centering
    \caption{\small\textbf{\added{Mann-Whitney U test results of \sys against \aflpp, \aflppminset, \aflppcq, and \aflppcmin on FuzzBench targets for 24 hours over 10 runs. Statistically significant results ($p < 0.05$) are in bold.}}}
    \begin{tabular}{lrrrr}
    \toprule
    \textbf{Targets} & \textbf{\aflpp} & \textbf{\aflppminset} & \textbf{\aflppcq} & \textbf{\aflppcmin} \\
    \midrule
    bloaty      & \textbf{1.01e-03} & \textbf{1.73e-02} & \textbf{1.83e-04} & \textbf{1.83e-04} \\
    curl        & \textbf{1.73e-02} & \textbf{1.71e-03} & \textbf{1.83e-04} & \textbf{7.69e-04} \\
    harfbuzz    & \textbf{3.28e-04} & \textbf{1.82e-04} & \textbf{1.82e-04} & \textbf{1.82e-04} \\
    jsoncpp     & \textbf{1.46e-03} & \textbf{2.38e-04} & \textbf{1.62e-03} & \textbf{1.83e-02} \\
    lcms        & 5.96e-01          & 6.38e-02          & \textbf{7.62e-04} & 5.85e-02 \\
    libjpeg     & \textbf{3.61e-03} & \textbf{2.16e-03} & \textbf{2.45e-04} & \textbf{4.37e-04} \\
    libpcap     & 4.08e-01          & 1.00e+00          & 1.44e-01          & 1.44e-01 \\
    libpng      & 1.94e-01          & \textbf{5.31e-04} & \textbf{2.11e-04} & \textbf{4.41e-04} \\
    libxml      & \textbf{1.83e-04} & \textbf{1.83e-04} & \textbf{1.83e-04} & \textbf{1.83e-04} \\
    libxslt     & 1.98e-01          & 6.23e-01          & \textbf{1.01e-03} & \textbf{5.80e-03} \\
    openh264    & 6.50e-01          & 6.40e-02          & 4.50e-01          & \textbf{1.83e-04} \\
    openssl     & \textbf{1.83e-04} & \textbf{1.82e-04} & \textbf{1.83e-04} & \textbf{1.82e-04} \\
    openthread  & \textbf{7.26e-03} & \textbf{1.82e-04} & \textbf{1.83e-04} & \textbf{2.46e-04} \\
    re2         & 5.45e-01          & 8.80e-01          & \textbf{1.80e-04} & \textbf{1.81e-04} \\
    sqlite3     & 4.95e-01          & \textbf{4.57e-03} & \textbf{1.81e-04} & \textbf{2.82e-03} \\
    stbi        & 4.73e-01          & 4.96e-01          & \textbf{1.13e-02} & \textbf{7.29e-03} \\
    vorbis      & \textbf{1.37e-02} & \textbf{1.37e-02} & \textbf{1.71e-02} & \textbf{1.88e-02} \\
    woff2       & \textbf{3.74e-02} & \textbf{4.53e-03} & \textbf{1.81e-04} & \textbf{1.69e-03} \\
    zlib        & 3.75e-01          & \textbf{2.81e-02} & 6.89e-01          & 1.47e-01 \\
    \bottomrule
    \end{tabular}
    \label{tab:fuzzbench_mwu}
\end{table}

\begin{table}[!ht]
\small
    \centering
    \caption{\small\textbf{\added{Mann-Whitney U test results of \sys against \aflpp, \aflppminset, \aflppcq, and \aflppcmin on standalone programs for 24 hours over 10 runs. Statistically significant results ($p < 0.05$) are in bold.}}}
    \begin{tabular}{lrrrr}
    \toprule
    \textbf{Targets} & \textbf{\aflpp} & \textbf{\aflppminset} & \textbf{\aflppcq} & \textbf{\aflppcmin} \\
    \midrule
    bsdtar      & \textbf{5.83e-04} & \textbf{1.83e-04} & \textbf{1.83e-04} & 5.71e-01 \\
    exiv2       & \textbf{2.82e-03} & \textbf{7.62e-04} & \textbf{1.13e-02} & 3.07e-01 \\
    ffmpeg      & \textbf{1.83e-04} & \textbf{1.83e-04} & \textbf{1.83e-04} & \textbf{2.46e-04} \\
    jhead       & 1.00e+00          & 1.00e+00          & 3.68e-01          & 7.79e-02 \\
    jq          & \textbf{5.83e-04} & \textbf{1.82e-04} & \textbf{1.83e-04} & \textbf{3.12e-02} \\
    mp3gain     & \textbf{3.38e-02} & \textbf{8.65e-04} & \textbf{1.80e-04} & \textbf{1.81e-04} \\
    mp42aac     & \textbf{7.29e-03} & \textbf{3.30e-04} & \textbf{1.83e-04} & \textbf{2.11e-02} \\
    nm-new      & \textbf{3.30e-04} & \textbf{1.83e-04} & \textbf{1.83e-04} & \textbf{3.61e-03} \\
    objdump     & \textbf{1.71e-03} & \textbf{1.83e-04} & \textbf{1.83e-04} & \textbf{5.80e-03} \\
    pdftotext   & \textbf{1.83e-04} & \textbf{1.82e-04} & \textbf{1.83e-04} & \textbf{1.71e-03} \\
    readelf     & 5.39e-02          & \textbf{1.01e-03} & \textbf{1.83e-04} & \textbf{1.83e-04} \\
    size        & 2.41e-01          & \textbf{1.83e-04} & \textbf{1.83e-04} & \textbf{1.83e-04} \\
    strip-new   & 3.64e-01          & \textbf{1.83e-04} & \textbf{1.83e-04} & \textbf{1.83e-04} \\
    tcpdump     & 1.04e-01          & \textbf{7.69e-04} & \textbf{1.83e-04} & \textbf{1.83e-04} \\
    xmllint     & \textbf{1.70e-03} & \textbf{1.83e-04} & \textbf{3.30e-04} & 6.78e-01 \\
    \bottomrule
    \end{tabular}
    \label{tab:standalone_mwu}
\end{table}

\begin{table*}[!t]
\small
\centering
\caption{\small\textbf{Mann--Whitney U test results (p-values) of \sys against the corresponding baselines for libAFL and Centipede integrations on FuzzBench targets (24 hours, 10 runs). Statistically significant results (\(p < 0.05\)) are in bold.}}
\label{tab:libafl_centipede_mwu}
\begin{tabular}{l r @{\hspace{1.6em}} l r}
\toprule

\multicolumn{4}{c}{\textbf{libAFL Based}} \\
\midrule
\textbf{Targets} & \textbf{Results} &
\textbf{Targets} & \textbf{Results} \\
\midrule
bloaty     & \textbf{3.28e-04} & openthread & 3.26e-01 \\
curl       & /                 & re2        & \textbf{2.78e-04} \\
harfbuzz   & \textbf{1.60e-02} & zlib       & 7.33e-01 \\
jsoncpp    & 4.72e-01          & libpcap    & 3.68e-01 \\
lcms       & 3.46e-01          & libjpeg    & \textbf{7.03e-04} \\
openh264   & /                 & libpng     & 9.02e-01 \\
openssl    & /                 & libxml     & \textbf{3.05e-02} \\
libxslt    & \textbf{1.26e-03} & sqlite3    & 5.40e-01 \\
stbi       & \textbf{3.42e-02} & vorbis     & 7.57e-02 \\
woff2      & 1.81e-01          &            &          \\

\midrule
\multicolumn{4}{c}{\textbf{Centipede Based}} \\
\midrule
\textbf{Targets} & \textbf{Results} &
\textbf{Targets} & \textbf{Results} \\
\midrule
bloaty     & \textbf{3.61e-03} & openthread & \textbf{2.20e-03} \\
curl       &       /            & re2        & \textbf{1.70e-03} \\
harfbuzz   & \textbf{1.71e-03} & zlib       & 5.95e-01 \\
jsoncpp    & 1.00e+00          & libpcap    & 6.39e-02 \\
lcms       & \textbf{1.55e-02} & libjpeg    & \textbf{2.11e-02} \\
openh264   & \textbf{4.57e-03} & libpng     & 1.00e+00 \\
openssl    & \textbf{4.48e-02} & libxml     & 1.21e-01 \\
libxslt    & \textbf{7.69e-04} & sqlite3    & 1.00e+00 \\
stbi       & \textbf{5.80e-04} & vorbis     & 4.96e-01 \\
woff2      & 1.86e-01          &            &          \\

\bottomrule
\end{tabular}
\end{table*}

\clearpage

\bibliographystyle{ACM-Reference-Format}

\bibliography{paper}

\end{document}